\documentclass[aps,prl,reprint,longbibliography]{revtex4-1}

\usepackage{amsmath,amsfonts,bm,graphicx,hyperref,verbatim,mathrsfs,units}

\usepackage{color,colordvi}
\definecolor{blue}{rgb}{0,0,1}
\definecolor{grey}{rgb}{0.6,0.6,0.6}

\begin{document}


\title{Certifying Non-Classical Behavior for Negative Keldysh Quasi-Probabilities}

\author{Patrick P. Potts}
\email{patrick.hofer@teorfys.lu.se}
\thanks{The author was previously known as Patrick P. Hofer.}
\affiliation{Physics Department and NanoLund, Lund University, Box 118,  22100 Lund, Sweden.}

\date{\today}

\begin{abstract}
	We introduce an experimental test for ruling out classical explanations for the statistics obtained when measuring arbitrary observables at arbitrary times using individual detectors. This test requires some trust in the measurements, represented by a few natural assumptions on the detectors. In quantum theory, the considered scenarios are well captured by von Neumann measurements. These can be described naturally in terms of the Keldysh quasi-probability distribution (KQPD), and the imprecision and backaction exerted by the measurement apparatus. We find that classical descriptions can be ruled out from measured data if and only if the KQPD exhibits negative values. We provide examples based on simulated data, considering the influence of a finite amount of statistics. In addition to providing an experimental tool for certifying non-classicality, our results bestow an operational meaning upon the non-classical nature of negative quasi-probability distributions such as the Wigner function and the full counting statistics.
\end{abstract}


\maketitle


\textit{Introduction.---} The theory of quantum mechanics contains ingredients that are absent in classical theories, such as entanglement, wave-function collapse, and superposition of arbitrary states \cite{haroche:1998,parisi:book,dowling:2003}. In some scenarios, these ingredients are beneficial (e.g., quantum information \cite{nielsen:book}), while in other scenarios, they provide limitations (e.g., quantum noise in measurement and amplification \cite{clerk:rmp}). The realm of possibilities that are enabled or prohibited by quantum mechanics is a highly non-trivial subject of current research. 

At the heart of this problem lies the question: \textit{``which observations cannot be explained by classical theories?''} A strong result in this direction is provided by Bell inequalities \cite{bell:1964}. With the help of such inequalities, observed data alone can rule out any theory that fulfills a natural definition of locality \cite{brunner:rmp}. While this is an extremely powerful result, locality is a rather specific requirement and does not encompass all classical theories \cite{newton}.

Another well established approach for testing for non-classicality is given by the Glauber-Sudarshan $P$-function in quantum optics \cite{glauber:1963,sudarshan:1963}. If a state is described by a $P$-function that cannot be interpreted as a probability distribution, then some measurable intensity correlators resulting from this state cannot be described by classical electrodynamics \cite{mandel:1986,vogel:2008}. In contrast to Bell inequalities, the measurement device thus has to be trusted to produce intensity correlators of light.
	
Arguably the most striking difference to classical theories is the fact that observables cannot be described using positive probability distributions in quantum mechanics. Leggett-Garg inequalities \cite{leggett:1985} provide a test for non-classicality based on this criterion. However, an additional assumption of non-invasive measurability which is not generally justified complicates the conclusions \cite{emary:2014}.

\begin{figure}
	\centering
	\includegraphics[width=\columnwidth]{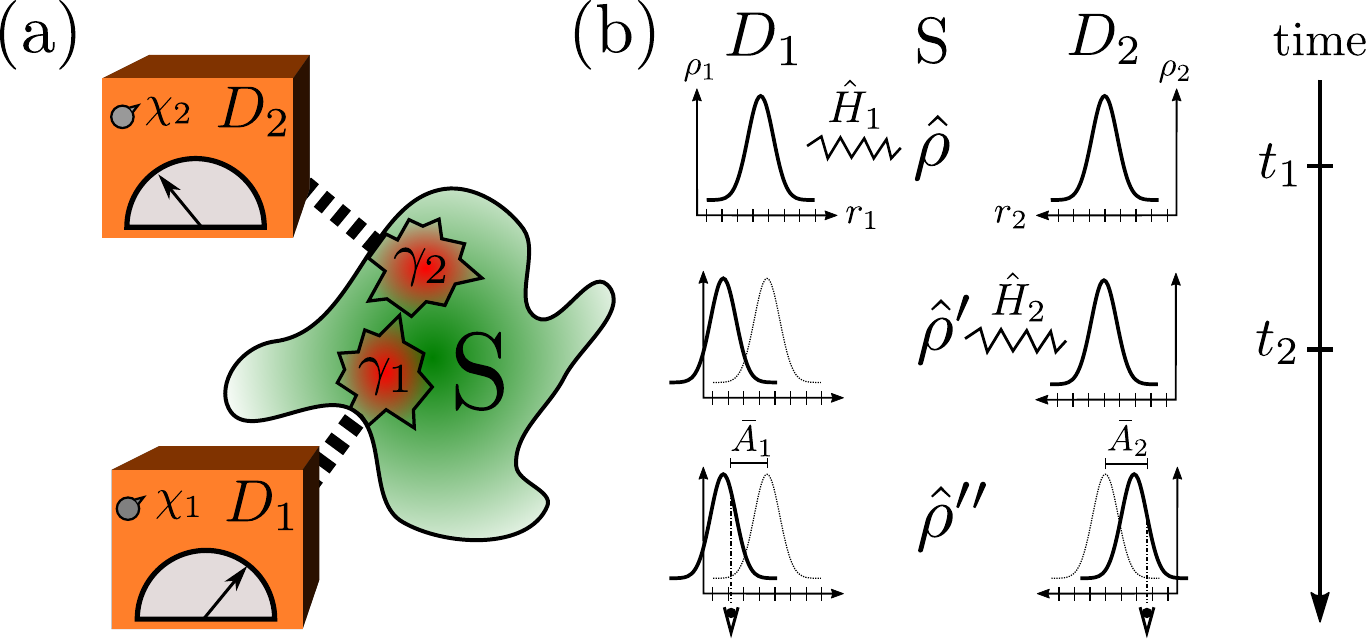}
	\caption{(a) Sketch of the setup. Two observables are measured by detectors $(D_j)$ coupled to the system ($S$). The detectors come with a knob $(\chi_j)$ and disturb the system ($\gamma_j$). (b) Illustration of von Neumann measurements. Detectors are quantum mechanical systems that couple to the system of interest at times $t_j$ via the Hamiltonian $\hat{H}_j$. The interaction shifts the probability distribution $\rho_j$ of the detectors by an amount depending on the system state $\hat{\rho}$. After the interaction, a projective position measurement is performed on the detectors resulting in the outcome $\bar{A}_j$.}
	\label{fig:model}
\end{figure}

In this letter, we we provide a test for non-classicality which rules out any description based on positive probabilities under a few realistic assumptions on the measurement apparatus.
To this end, we consider scenarios where observables are measured using individual detectors, see Fig.~\ref{fig:model}. In quantum theory, such scenarios are well described by von Neumann type measurements \cite{vonneumann:book,stenholm:1992,busch:2009,clerk:2010}, where observables of interest are coupled to detectors which are subsequently measured projectively. The probability distribution describing the measurement outcomes have a natural description in terms of a quasi-probability distribution that we abbreviate with KQPD due to its reminiscence of the Keldysh path-integral formulation \cite{hofer:2017,aron:2018}. The KQPD depends on the observables of interest and can reduce to the Wigner function \cite{qm_phase:book} or the full counting statistics \cite{nazarov:book2}. Other applications include quantum thermodynamics \cite{esposito:2009,solinas:2015,solinas:2016,baumer:2018,chiara:2018,lostaglio:2018}, quantum optics \cite{clerk:2011}, generalized Wigner functions \cite{schwonnek:2018}, weak values \cite{hofer:2017} (see also \cite{hallaji:2017,sinclair:2018,halpern:2018}), and non-equilibrium phenomena in quantum systems \cite{aron:2018}. Importantly, the KQPD can become negative, indicating non-classical behavior \cite{belzig:2001,kenfack:2004,deleglise:2008,bednorz:2010,clerk:2010,clerk:2011,bednorz:2012,hofer:2016}. Here we put this non-classical feature on a firmer footing by taking an operational approach. To this end, we put forward a classical model for measurements based on individual detectors. This model is based on a few natural assumptions on the detectors and results in an experimentally accessible inequality. We show that within quantum theory, negativity in the KQPD is a necessary and sufficient condition to violate the inequality, ruling out a classical description. Just like negativity in the $P$-function rules out an explanation by classical electrodynamics (as long as the detectors can be trusted to produce intensity correlators), negativity in the KQPD rules out an explanation based on positive probabilities, as long as the measurement apparatus can be trusted to fulfill the assumptions specified below.

In contrast to Leggett-Garg inequalities, non-invasiveness of the measurement is not required. The proposed experimental test of non-classical behavior is therefore not subject to a \textit{clumsiness} \cite{emary:2014} or a \textit{finite precision} loophole \cite{kent:1999,meyer:1999}. The model is not necessarily local or non-contextual \cite{spekkens:2008,cabello:2008,kirchmair:2009}.

{Before we introduce the classical model, we provide the quantum mechanical (QM) description of the scenario under investigation, sketched in Fig.~\ref{fig:model}. While we assume this to be the correct description, we stress that our test for non-classicality does not rely on the QM-model.}

\textit{The KQPD.---} The QM-model relies on the KQPD which is discussed in detail in Ref.~\cite{hofer:2017}. It encodes the joint fluctuations of multiple observables of interest. For simplicity, we consider the situation where we are interested in two observables $\hat{A}_1$ and $\hat{A}_2$ at times $t_1$ and $t_2$ respectively. The generalization to more observables is straightforward. Let us further consider the situation where $t_2$ either comes immediately after $t_1$ (subsequent measurements) or where $t_1=t_2$ (simultaneous measurements). The KQPD is then defined as ($\hbar=1$)
\begin{equation}
\label{eq:kqpd}
\mathcal{P}(\boldsymbol{A}|\boldsymbol{\gamma}) =\int  \frac{d\boldsymbol{\lambda}}{(2\pi)^2}e^{i\boldsymbol{\lambda}\cdot\boldsymbol{A}}{\rm Tr}\left\{\hat{Q}(\boldsymbol{\lambda},\boldsymbol{\gamma})\hat{\rho}\hat{Q}^\dagger(-\boldsymbol{\lambda},\boldsymbol{\gamma})\right\},
\end{equation}
where $\hat{Q}=\exp[-i\left(\frac{\lambda_2}{2}+\gamma_2\right)\hat{A}_2]\exp[-i\left(\frac{\lambda_1}{2}+\gamma_1\right)\hat{A}_1]$ for subsequent and $\hat{Q}=\exp[-i\sum_{j=1,2}\left({\lambda_j}/2+\gamma_j\right)\hat{A}_j]$ for simultaneous measurements. The state before the measurement is denoted by $\hat{\rho}$. We grouped the observables into a vector $\boldsymbol{A}=(A_1,A_2)$ and similarly for $\boldsymbol{\lambda}$ and $\boldsymbol{\gamma}$. As shown below, the variables $\gamma_j$ are necessary to take into account the backaction exerted by the measurement and can be seen as random variables determined by the detectors. A physical motivation for the definition in Eq.~\eqref{eq:kqpd} is provided below, by Eq.~\eqref{eq:measqm}.

If $[\hat{A}_1,\hat{A}_2]\neq 0$, the measurement of $\hat{A}_1$ may influence the measurement of $\hat{A}_2$ and a description of the system in terms of pre-determined values of $A_1$ and $A_2$ is not generally possible. In this case, the KQPD may become negative. It has been shown that such negativity requires the system to be in a superposition of states that correspond to different values for the observable $A_1$ \cite{hofer:2016}. Negativity in the KQPD can thus be seen as an indicator for non-classical behavior. However, in an experiment, the negativity of the KQPD is masked by measurement imprecision and backaction, rendering the measured probability distribution strictly non-negative. The inequality that we introduce below relies on a way to unmask the KQPD experimentally.

\textit{The QM-model.---} We consider two detectors, one for each observable to be measured. The detectors can be described by canonically conjugate observables $\hat{r}_j$ and $\hat{\pi}_j$, and they are coupled to the system through the Hamiltonian \cite{vonneumann:book}
\begin{equation}
\label{eq:hammeas}
\hat{H}_j = \delta(t-t_j)\chi_j\hat{A}_j\hat{\pi}_j,
\end{equation}
where $j=1,2$, and $\chi_j$ denotes the measurement strength. We assume that the time-evolution induced by any Hamiltonian other than Eq.~\eqref{eq:hammeas} can be neglected during (and between) the measurements, noting that it is straightforward to include time-evolution between the measurements (for an investigation on detector memory effects, see Ref.~\cite{bulte:2018}). Equation \eqref{eq:hammeas} induces a displacement in the detector coordinates $\hat{r}_j$ which depends on the state of the system. After the interaction, a projective measurement of the detectors is performed to complete the measurement of the system observables $\hat{A}_j$. The measured distribution reads \cite{hofer:2017} (see also \cite{nazarov:2003,di_lorenzo:2011})
\begin{equation}
\label{eq:measqm}
P(\boldsymbol{A}|\boldsymbol{\chi})=\int d\boldsymbol{A}'d\boldsymbol{\gamma}\mathcal{P}(\boldsymbol{A}'|\boldsymbol{\gamma}) \prod_{j=1,2}\mathcal{W}_j(\bar{A}_j-\bar{A}'_j,\bar{\gamma}_j),
\end{equation}
where $\mathcal{W}_j(r,\pi)$ denotes the Wigner function of detector $j$ and we introduced $\bar{A}_j=\chi_jA_j$ and $\bar{\gamma}_j=\gamma_j/\chi_j$. This equation has a simple interpretation, motivating the definition in Eq.~\eqref{eq:kqpd}. The KQPD describes the intrinsic fluctuations of the observables, containing all the information of the system. These fluctuations are distorted by the measurement process, giving rise to the convolution with the Wigner functions of the detectors. The uncertainty in the position coordinates induces a fuzziness in the measurement (measurement imprecision) and the uncertainty in the momentum coordinates introduces a random kick in the measured observable through Eq.~\eqref{eq:hammeas} (measurement backaction). Due to the Heisenberg uncertainty relation, there exists a trade-off between imprecision and backaction \cite{clerk:rmp} which ensures that the measured distribution is always positive, even when the KQPD exhibits negativity. For an investigation of the classical limit of von Neumann type measurements, see Ref.~\cite{barnea:2017}.

\textit{The classical model.---} We now introduce a classical hidden-variable model that describes the situation sketched in Fig.~\ref{fig:model}\,(a). To this end, we assume that the system is described by a probability distribution $S(\boldsymbol{A}|\boldsymbol{\gamma})$. This distribution encodes the (hidden) values of the observables ($\boldsymbol{A}$) and takes into account that the presence of the detectors may modify the system behavior $(\boldsymbol{\gamma})$. The measured distribution can then be written in the completely general form
\begin{equation}
\label{eq:classgen}
P_{\rm cl}(\boldsymbol{A}|\boldsymbol{\chi}) = \int d\boldsymbol{A}'d\boldsymbol{\gamma} M(\boldsymbol{A},\boldsymbol{A}',\boldsymbol{\gamma}|\boldsymbol{\chi})S(\boldsymbol{A}'|\boldsymbol{\gamma}),
\end{equation}
where $\boldsymbol{\chi}$ describes the (changeable) detector settings. The function $M$ describes the effect of the detectors. We say that an observed probability distribution has a classical explanation if it can be described by the right-hand side of Eq.~\eqref{eq:classgen} with positive $S$ and $M$.

Equation \eqref{eq:classgen} is sufficiently general that it can essentially describe any observations. To rule out a classical explanation, we place some trust in the detectors and make the assumptions:
\begin{enumerate}
	\item \textbf{Uncorrelated detectors:} \begin{equation}
	\label{eq:ass1}
	M(\boldsymbol{A},\boldsymbol{A}',\boldsymbol{\gamma}|\boldsymbol{\chi})=\prod_j M_j({A}_j,{A}'_j,\gamma_j|\chi_j).\end{equation}
	\item \textbf{Uncorrelated imprecision and backaction:}
	\begin{equation}
	\label{eq:ass2} M_j({A}_j,{A}'_j,\gamma_j|\chi_j)=p_j(\gamma_j|\chi_j)D_j({A}_j,{A}'_j|\chi_j).\end{equation}
	\item \textbf{Backaction only affects the other observable:}
	\begin{equation}
	\label{eq:ass3} \int dA_{k}S(\boldsymbol{A}|\gamma_j,\gamma_k=0)\equiv S({A}_j|\gamma_j)=S(A_j).\end{equation}
	\item \textbf{Translational invariance:}
	\begin{equation}
	\label{eq:ass4} D_j({A}_j,{A}'_j|\chi_j)=D_j({A}_j-{A}'_j|\chi_j).\end{equation}
	\item \textbf{Detectors can be detached:}
	\begin{equation}
	\label{eq:ass5} \lim_{\chi_j\rightarrow0}p_j(\gamma_j|\chi_j)D_j({A}_j-{A}'_j|\chi_j)=\delta(\gamma_j)U(A_j).\end{equation}
	
\end{enumerate}
In the spirit of the considered scenario, the first assumption allows us to treat the detectors as individual objects (note that this assumption is also present in the Bell scenario). Assumptions 2 and 3 ensure that the backaction of a detector does not interfere with its own measurement, i.e., a detector's output is independent of its backaction on the system. In Eq.~\eqref{eq:ass3}, we introduced the distribution relevant for measuring a single variable, $S(A_j)$, which is assumed to be independent of the backaction of its own detector. In assumption 5, $U$ denotes the uniform distribution and we defined $\gamma_j=0$ to denote the absence of any backaction of detector $j$. We note that our assumptions only include the effect of the detectors. On a qualitative level, one can thus replace our assumptions with the notion of having control over measurements of single observables and preventing any cross-talk between the detectors.


\textit{Certifying non-classicality.---} We denote by $P(A_j|\chi_j)$ the distribution that describes a measurement of a single observable. We further denote the Fourier transform of any distribution with a tilde $\tilde{P}(\lambda)=\int dA\exp(-i\lambda A)P(A)$. We then consider the quantity
\begin{equation}
\label{eq:reckqpd}
K = \frac{1}{(2\pi)^2}\int d\boldsymbol{\lambda}e^{i\boldsymbol{\lambda}\cdot\boldsymbol{A}}\tilde{P}(\boldsymbol{\lambda}|\boldsymbol{\chi})\prod_{j=1,2}\frac{\tilde{P}(\lambda_j|\chi_j')}{\tilde{P}(\lambda_j|\chi_j)},
\end{equation}
where we note that the right-hand side only contains Fourier transforms of \textit{measurable} probability distributions. If the measurement is described by our classical model, we can write this quantity as \cite{supp}
\begin{equation}
\label{eq:reccl}
K_{\rm cl} = \int d\boldsymbol{A}'d\boldsymbol{\gamma}S(\boldsymbol{A}'|\boldsymbol{\gamma}) \prod_{j=1,2}p_j(\gamma_j|\chi_j)D_j({A}_j-{A}'_j|\chi_j').
\end{equation}
This equation is very similar to Eq.~\eqref{eq:classgen} (under our assumptions) with the only difference that $\chi_j$ is replaced by $\chi_j'$ in the measurement imprecision term $D_j$. Within our assumptions, the measurement imprecision of the detectors can be corrected for. We end up with a distribution where the backaction is determined by $\chi_j$ and the imprecision by $\chi_j'$. In our classical model, this still results in a positive distribution
\begin{equation}
\label{eq:inequality}
K_{\rm cl}\geq 0.
\end{equation}
Any violation of this inequality implies that the observed data cannot be explained by Eq.~\eqref{eq:classgen} with positive $S$ and $M$ that satisfy the five assumptions. Trusting the detectors (i.e., the assumptions) then allows us to conclude that no explanation in terms of positive probabilities is possible. The assumptions thus introduce loopholes since a violation of Eq.~\eqref{eq:inequality} could in principle result from their breakdown.

In quantum mechanics, the delicate interplay between backaction and imprecision is what masks the negativity of the KQPD. This may result in a violation of the inequality. Using detectors with positive Wigner functions that factorize in a position and a momentum part ensures that our assumptions on the detectors are satisfied. The quantity $K$ is then given by an expression analogous to Eq.~\eqref{eq:reccl}, with $S$ replaced by the KQPD $\mathcal{P}$. This can be seen by plugging Eq.~\eqref{eq:measqm}, and a similar expression for single observables, into Eq.~\eqref{eq:reckqpd}. A positive KQPD then immediately ensures $K\geq0$. In the limit where $\chi_j\rightarrow0$ and $\chi_j'\rightarrow\infty$, we find $K\rightarrow\mathcal{P}$. Whenever the KQPD exhibits negativity, we can thus find $K<0$, violating the inequality in Eq.~\eqref{eq:inequality}. Since the assumptions on the detectors are met, this implies that the measured data cannot be explained by positive probability distributions. Negativity in the KQPD is therefore a necessary and sufficient condition for certifying non-classicality.

\begin{figure*}
	\centering
	\includegraphics[width=\textwidth]{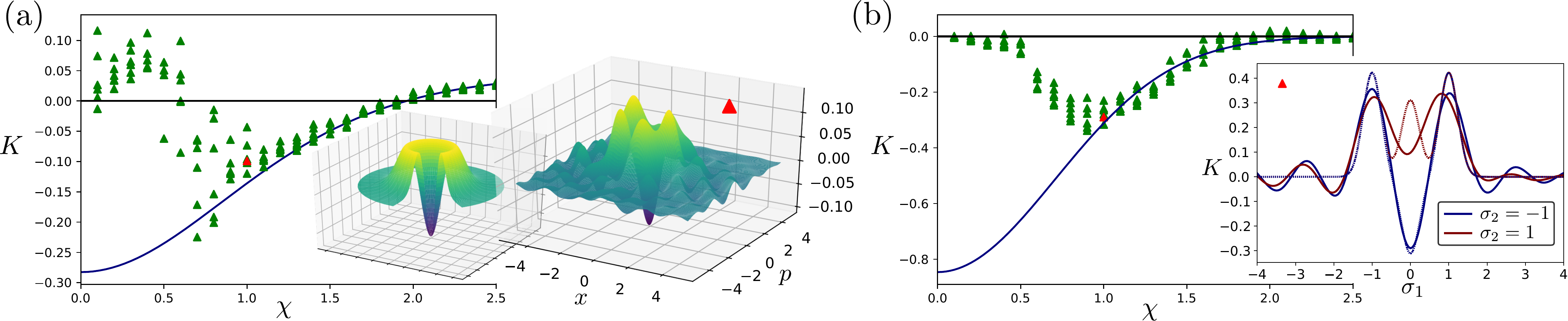}
	\caption{Certifying non-classicality. (a) Simultaneous measurement of both quadratures in a single-mode Fock state containing one photon. (b) Two subsequent spin measurements in different directions on a spin one-half particle. The large panels show $K$ for values $\boldsymbol{A}$ that maximize the negativity [$x=0$, $p=0$ for (a), $\sigma_1=0$, $\sigma_2=-1$ for (b)]. The solid line corresponds to the exact value of $K$ [Eq.~\eqref{eq:reckqpd}], the triangles to the estimate $K_{\rm est}$ based on numerical simulations [Eq.~\eqref{eq:kest}]. The side panels show the full estimate of $K$ for a single data point. In (a), the small side-panel shows the exact distribution $K$. In (b), the dashed lines correspond to the exact $K$. As the measurement strength $\chi$ increases, the estimate becomes more reliable but the backaction decreases the negativity in $K$. The simulations are based on $15\,000$ individual measurements of the observables and $30\,000$ joint measurements. Other parameters: (a) $\chi'=5$, $c_o=0.011$, $\lambda_c=10$. (b)  $\chi'=3$, $c_o=0.01$, $\lambda_c=12$.}
	\label{fig:examples}
\end{figure*}

\textit{Examples.---} We now illustrate how our classical model can be ruled out from experimental (in our case, simulated) data by violating the inequality in Eq.~\eqref{eq:inequality}. We consider two examples: The simultaneous measurement of position and momentum, and two subsequent, non-commuting Stern-Gerlach type spin measurements. For both examples, we consider identical detectors that are described by the Wigner function (throughout, we consider dimensionless units for position and momentum)
\begin{equation}
\label{eq:wignerdet}
 \mathcal{W}_j(r_j,p_j)=\frac{1}{\pi}e^{-(r_j^2+p_j^2)}/\pi,
\end{equation}
 corresponding to unsqueezed Gaussian states of minimal uncertainty. As demanded by assumption 2, they factorize into distributions for position (imprecision) and momentum (backaction).

We first consider a simultaneous measurement of position and momentum on a single-photon Fock state described by the Wigner function
\begin{equation}
\label{eq:wignerfock}
\mathcal{W}(x,p)=\frac{1}{\pi}\left[2(x^2+p^2)-1\right]e^{-(x^2+p^2)}.
\end{equation}
In this case, our quantum mechanical model reduces to the Arthurs-Kelly model \cite{arthurs:1964}. We note that such a measurement can be implemented by heterodyne detection \cite{leonhardt:1995}, see Ref.~\cite{eichler:2011} for an experimental realization.
As discussed in detail in Ref.~\cite{hofer:2017}, the KQPD for the simultaneous position and momentum measurement is given by $\mathcal{W}(x-\gamma_p/2,p+\gamma_x/2)$. Choosing equal measurement strengths $\chi_x=\chi_p=\chi$ and $\chi_x'=\chi_p'=\chi'$ we then find (see supplemental information for details \cite{supp})
\begin{equation}
\label{eq:kfock}
K = \frac{1}{\pi(1+g)^3}e^{-\frac{(x^2+p^2)}{1+g}}\left[2(x^2+p^2)-1+g^2\right],
\end{equation}
where $g=(\chi/2)^2+1/(\chi')^2$. We note that in the limit $\chi\rightarrow0$ and $\chi'\rightarrow\infty$, we have $g\rightarrow0$ and Eq.~\eqref{eq:kfock} reduces to Eq.~\eqref{eq:wignerfock}. As long as $g<1$, we find $K<0$ at the origin, see Fig.~\ref{fig:examples}\,(a).

Equation \eqref{eq:kfock} implies that the smaller $\chi$, the stronger the negativity in the measurable quantity $K$. Weaker measurements thus always seem to be preferable. This is only true under the assumption that $K$ can be estimated precisely. Strictly speaking, this requires an infinite amount of data. For a finite and fixed number of measurements, we will find a trade-off between having large negative values in $K$ (requiring small $\chi$) and being able to reliably estimate $K$ (requiring large $\chi$). To estimate $K$, we consider an experiment with $N$ measurements resulting in outcomes $x_j$. We define the empirical characteristic function \cite{feuerverger:1977}
	\begin{equation}
	\label{eq:empchar}
	Y_\lambda = \frac{1}{N}\sum_{j=1}^{N}e^{-i\lambda x_j},
	\end{equation}
which provides an unbiased estimator of the characteristic function (i.e., the Fourier transform of the probability distribution). We note that it is imprecise for large values of $\lambda$, where the characteristic function is a small number.
For $K$, we introduce the estimator
\begin{equation}
\label{eq:kest}
K_{\rm est} = \begin{cases}
\int_{-\lambda_c}^{\lambda_c} \frac{d\boldsymbol{\lambda}}{(2\pi)^2}e^{i\boldsymbol{\lambda}\cdot\boldsymbol{A}}Y_{\boldsymbol{\lambda}}\frac{Y_{\lambda_x}'}{Y_{\lambda_x}}\frac{Y_{\lambda_p}'}{Y_{\lambda_p}} \hspace{.2cm}{\rm for}\hspace{0.2cm}|Y_{\lambda_{x/p}}| > c_o,\\
0 \hspace{.5cm}{\rm otherwise},
\end{cases}
\end{equation}
where $\boldsymbol{\lambda}\cdot\boldsymbol{A}=\lambda_x x+\lambda_p p$.
Here the different empirical characteristic functions are labeled by $\boldsymbol{\lambda}$ for the joint measurement and by a prime for the measurements with strength $\chi'$. Two empirical cut-offs increase the stability of the estimator. The first, $c_o$, ensures that values of $\boldsymbol{\lambda}$ where we divide by a very small number are not taken into account. The second, $\lambda_c$, allows for integrating over a finite domain. The estimator in Eq.~\eqref{eq:kest} is illustrated in Fig.~\ref{fig:examples}\,(a) for simulated data. For large values of $\chi$, it is both accurate and precise. As $\chi$ becomes smaller, the spread of the estimates increases (the precision is reduced). Eventually, the cut-off $c_o$ prevents an accurate estimation because the true characteristic function becomes very small for almost all values of $\lambda_{x/p}$. As expected, we find a trade-off between large $\chi$, where the negativity in $K$ is not very pronounced, and small $\chi$, where it is hard to estimate $K$.

Our second example is provided by subsequent, non-commuting measurements on a two-level system (for a recent experimental implementation of non-commuting spin measurements, see Ref.~\cite{hacohen:2016}, for a detailed discussion on simultaneous spin measurements, see Ref.~\cite{perarnau:2017}). We consider the system to be in a pure state $|+\rangle$, which is an eigenstate of the Pauli matrix $\hat{\sigma}_x$. We then make a measurement of $\hat{\sigma}_z$ with strength $\chi_1=\chi$, followed by a projective measurement of $\hat{\sigma}_x$. The KQPD for this system is discussed in Ref.~\cite{hofer:2017} and given in the supplemental information \cite{supp}. Because it is unavoidable that the first measurement influences the second one, the KQPD exhibits negativity. Since the second measurement is projective, we only correct for the measurement imprecision of the first measurement, choosing $\chi_2=\chi_2'\rightarrow \infty$ in Eq.~\eqref{eq:reckqpd}. All distributions can then be given as densities in the continuous variable $\sigma_1$ and probabilities in the discrete variable $\sigma_2=\pm 1$. Certifying non-classicality of this system is illustrated in Fig.~\ref{fig:examples}\,(b), where we show both $K$ as well as $K_{\rm est}$. We find the same qualitative results as for the simultaneous position and momentum measurement. The weaker the first measurement, the more pronounced the negativity but the less reliable is the estimate $K_{\rm est}$. Detailed calculations can be found in the supplemental information \cite{supp}.

\textit{Conclusions.---} We introduced a classical model for measurements that use individual detectors for different observables. Under five natural assumptions, we find the inequality $K\geq0$. Any violation of this inequality implies that either no description in terms of positive probabilities is possible, or one of the assumptions on the detectors is not met. In scenarios which are well described by quantum mechanical von Neumann measurements, we find that $K$ can become negative if and only if the KQPD exhibits negative values. In this case, $K$ provides a way of approximating the KQPD from measurable probability distributions.
This is possible because measurement imprecision is a property of the detector alone and can thus be inferred and corrected for. In weak measurements, where backaction becomes small, correcting for the measurement imprecision ``unmasks'' the KQPD, exposing its negativity.

Our classical model is appropriate whenever individual detectors are used to measure different observables. The introduced operational procedure for certifying non-classicality is thus of broad experimental relevance and it puts the non-classical nature of the negative values in the KQPD on a firmer footing.

\textit{Acknowledgements.---} I acknowledge fruitful discussions with M.-O. Renou, N. Brunner, and P. Samuelsson. I acknowledge funding from the European Union's Horizon 2020 research and innovation programme under the Marie Sk{\l}odowska-Curie Grant Agreement No. 796700. This work was supported by the Swedish Research Council as well as the Swiss National Science Foundation.

\bibliography{biblio}

\begin{thebibliography}{55}%
\makeatletter
\providecommand \@ifxundefined [1]{%
 \@ifx{#1\undefined}
}%
\providecommand \@ifnum [1]{%
 \ifnum #1\expandafter \@firstoftwo
 \else \expandafter \@secondoftwo
 \fi
}%
\providecommand \@ifx [1]{%
 \ifx #1\expandafter \@firstoftwo
 \else \expandafter \@secondoftwo
 \fi
}%
\providecommand \natexlab [1]{#1}%
\providecommand \enquote  [1]{``#1''}%
\providecommand \bibnamefont  [1]{#1}%
\providecommand \bibfnamefont [1]{#1}%
\providecommand \citenamefont [1]{#1}%
\providecommand \href@noop [0]{\@secondoftwo}%
\providecommand \href [0]{\begingroup \@sanitize@url \@href}%
\providecommand \@href[1]{\@@startlink{#1}\@@href}%
\providecommand \@@href[1]{\endgroup#1\@@endlink}%
\providecommand \@sanitize@url [0]{\catcode `\\12\catcode `\$12\catcode
  `\&12\catcode `\#12\catcode `\^12\catcode `\_12\catcode `\%12\relax}%
\providecommand \@@startlink[1]{}%
\providecommand \@@endlink[0]{}%
\providecommand \url  [0]{\begingroup\@sanitize@url \@url }%
\providecommand \@url [1]{\endgroup\@href {#1}{\urlprefix }}%
\providecommand \urlprefix  [0]{URL }%
\providecommand \Eprint [0]{\href }%
\providecommand \doibase [0]{http://dx.doi.org/}%
\providecommand \selectlanguage [0]{\@gobble}%
\providecommand \bibinfo  [0]{\@secondoftwo}%
\providecommand \bibfield  [0]{\@secondoftwo}%
\providecommand \translation [1]{[#1]}%
\providecommand \BibitemOpen [0]{}%
\providecommand \bibitemStop [0]{}%
\providecommand \bibitemNoStop [0]{.\EOS\space}%
\providecommand \EOS [0]{\spacefactor3000\relax}%
\providecommand \BibitemShut  [1]{\csname bibitem#1\endcsname}%
\let\auto@bib@innerbib\@empty
\bibitem [{\citenamefont {Haroche}(1998)}]{haroche:1998}%
  \BibitemOpen
  \bibfield  {author} {\bibinfo {author} {\bibfnamefont {S.}~\bibnamefont
  {Haroche}},\ }\bibfield  {title} {\enquote {\bibinfo {title} {Entanglement,
  decoherence and the quantum/classical boundary},}\ }\href@noop {} {\bibfield
  {journal} {\bibinfo  {journal} {Phys. Today}\ }\textbf {\bibinfo {volume}
  {51}},\ \bibinfo {pages} {36} (\bibinfo {year} {1998})}\BibitemShut {NoStop}%
\bibitem [{\citenamefont {Parisi}\ and\ \citenamefont
  {Auletta}(2000)}]{parisi:book}%
  \BibitemOpen
  \bibfield  {author} {\bibinfo {author} {\bibfnamefont {G.}~\bibnamefont
  {Parisi}}\ and\ \bibinfo {author} {\bibfnamefont {G.}~\bibnamefont
  {Auletta}},\ }\href {\doibase 10.1142/4194} {\emph {\bibinfo {title}
  {Foundations and Interpretation of Quantum Mechanics}}}\ (\bibinfo
  {publisher} {World Scientific},\ \bibinfo {year} {2000})\BibitemShut
  {NoStop}%
\bibitem [{\citenamefont {Dowling}\ and\ \citenamefont
  {Milburn}(2003)}]{dowling:2003}%
  \BibitemOpen
  \bibfield  {author} {\bibinfo {author} {\bibfnamefont {J.~P.}\ \bibnamefont
  {Dowling}}\ and\ \bibinfo {author} {\bibfnamefont {G.~J.}\ \bibnamefont
  {Milburn}},\ }\bibfield  {title} {\enquote {\bibinfo {title} {Quantum
  technology: the second quantum revolution},}\ }\href {\doibase
  10.1098/rsta.2003.1227} {\bibfield  {journal} {\bibinfo  {journal} {Philos.
  Trans. Royal Soc. A}\ }\textbf {\bibinfo {volume} {361}},\ \bibinfo {pages}
  {1655} (\bibinfo {year} {2003})}\BibitemShut {NoStop}%
\bibitem [{\citenamefont {Nielsen}\ and\ \citenamefont
  {Chuang}(2010)}]{nielsen:book}%
  \BibitemOpen
  \bibfield  {author} {\bibinfo {author} {\bibfnamefont {M.~A.}\ \bibnamefont
  {Nielsen}}\ and\ \bibinfo {author} {\bibfnamefont {I.~L.}\ \bibnamefont
  {Chuang}},\ }\href {\doibase 10.1017/CBO9780511976667} {\emph {\bibinfo
  {title} {Quantum Computation and Quantum Information: 10th Anniversary
  Edition}}}\ (\bibinfo  {publisher} {Cambridge University Press},\ \bibinfo
  {year} {2010})\BibitemShut {NoStop}%
\bibitem [{\citenamefont {Clerk}\ \emph
  {et~al.}(2010{\natexlab{a}})\citenamefont {Clerk}, \citenamefont {Devoret},
  \citenamefont {Girvin}, \citenamefont {Marquardt},\ and\ \citenamefont
  {Schoelkopf}}]{clerk:rmp}%
  \BibitemOpen
  \bibfield  {author} {\bibinfo {author} {\bibfnamefont {A.~A.}\ \bibnamefont
  {Clerk}}, \bibinfo {author} {\bibfnamefont {M.~H.}\ \bibnamefont {Devoret}},
  \bibinfo {author} {\bibfnamefont {S.~M.}\ \bibnamefont {Girvin}}, \bibinfo
  {author} {\bibfnamefont {F.}~\bibnamefont {Marquardt}}, \ and\ \bibinfo
  {author} {\bibfnamefont {R.~J.}\ \bibnamefont {Schoelkopf}},\ }\bibfield
  {title} {\enquote {\bibinfo {title} {Introduction to quantum noise,
  measurement, and amplification},}\ }\href {\doibase
  10.1103/RevModPhys.82.1155} {\bibfield  {journal} {\bibinfo  {journal} {Rev.
  Mod. Phys.}\ }\textbf {\bibinfo {volume} {82}},\ \bibinfo {pages} {1155}
  (\bibinfo {year} {2010}{\natexlab{a}})}\BibitemShut {NoStop}%
\bibitem [{\citenamefont {Bell}(1964)}]{bell:1964}%
  \BibitemOpen
  \bibfield  {author} {\bibinfo {author} {\bibfnamefont {J.~S.}\ \bibnamefont
  {Bell}},\ }\bibfield  {title} {\enquote {\bibinfo {title} {On the {E}instein
  {P}odolsky {R}osen paradox},}\ }\href {\doibase
  10.1103/PhysicsPhysiqueFizika.1.195} {\bibfield  {journal} {\bibinfo
  {journal} {Physics}\ }\textbf {\bibinfo {volume} {1}},\ \bibinfo {pages}
  {195} (\bibinfo {year} {1964})}\BibitemShut {NoStop}%
\bibitem [{\citenamefont {Brunner}\ \emph {et~al.}(2014)\citenamefont
  {Brunner}, \citenamefont {Cavalcanti}, \citenamefont {Pironio}, \citenamefont
  {Scarani},\ and\ \citenamefont {Wehner}}]{brunner:rmp}%
  \BibitemOpen
  \bibfield  {author} {\bibinfo {author} {\bibfnamefont {N.}~\bibnamefont
  {Brunner}}, \bibinfo {author} {\bibfnamefont {D.}~\bibnamefont {Cavalcanti}},
  \bibinfo {author} {\bibfnamefont {S.}~\bibnamefont {Pironio}}, \bibinfo
  {author} {\bibfnamefont {V.}~\bibnamefont {Scarani}}, \ and\ \bibinfo
  {author} {\bibfnamefont {S.}~\bibnamefont {Wehner}},\ }\bibfield  {title}
  {\enquote {\bibinfo {title} {Bell nonlocality},}\ }\href {\doibase
  10.1103/RevModPhys.86.419} {\bibfield  {journal} {\bibinfo  {journal} {Rev.
  Mod. Phys.}\ }\textbf {\bibinfo {volume} {86}},\ \bibinfo {pages} {419}
  (\bibinfo {year} {2014})}\BibitemShut {NoStop}%
\bibitem [{new()}]{newton}%
  \BibitemOpen
  \href@noop {} {}\bibinfo {note} {Newton's theory of gravity is, for instance,
  non-local and can thus strictly speaking not be ruled out by a violation of a
  Bell inequality.}\BibitemShut {Stop}%
\bibitem [{\citenamefont {Glauber}(1963)}]{glauber:1963}%
  \BibitemOpen
  \bibfield  {author} {\bibinfo {author} {\bibfnamefont {Roy~J.}\ \bibnamefont
  {Glauber}},\ }\bibfield  {title} {\enquote {\bibinfo {title} {Coherent and
  incoherent states of the radiation field},}\ }\href {\doibase
  10.1103/PhysRev.131.2766} {\bibfield  {journal} {\bibinfo  {journal} {Phys.
  Rev.}\ }\textbf {\bibinfo {volume} {131}},\ \bibinfo {pages} {2766} (\bibinfo
  {year} {1963})}\BibitemShut {NoStop}%
\bibitem [{\citenamefont {Sudarshan}(1963)}]{sudarshan:1963}%
  \BibitemOpen
  \bibfield  {author} {\bibinfo {author} {\bibfnamefont {E.~C.~G.}\
  \bibnamefont {Sudarshan}},\ }\bibfield  {title} {\enquote {\bibinfo {title}
  {Equivalence of semiclassical and quantum mechanical descriptions of
  statistical light beams},}\ }\href {\doibase 10.1103/PhysRevLett.10.277}
  {\bibfield  {journal} {\bibinfo  {journal} {Phys. Rev. Lett.}\ }\textbf
  {\bibinfo {volume} {10}},\ \bibinfo {pages} {277} (\bibinfo {year}
  {1963})}\BibitemShut {NoStop}%
\bibitem [{\citenamefont {Mandel}(1986)}]{mandel:1986}%
  \BibitemOpen
  \bibfield  {author} {\bibinfo {author} {\bibfnamefont {L.}~\bibnamefont
  {Mandel}},\ }\bibfield  {title} {\enquote {\bibinfo {title} {Non-classical
  states of the electromagnetic field},}\ }\href {\doibase
  10.1088/0031-8949/1986/T12/005} {\bibfield  {journal} {\bibinfo  {journal}
  {Phys. Scr.}\ }\textbf {\bibinfo {volume} {1986}},\ \bibinfo {pages} {34}
  (\bibinfo {year} {1986})}\BibitemShut {NoStop}%
\bibitem [{\citenamefont {Vogel}(2008)}]{vogel:2008}%
  \BibitemOpen
  \bibfield  {author} {\bibinfo {author} {\bibfnamefont {W.}~\bibnamefont
  {Vogel}},\ }\bibfield  {title} {\enquote {\bibinfo {title} {Nonclassical
  correlation properties of radiation fields},}\ }\href {\doibase
  10.1103/PhysRevLett.100.013605} {\bibfield  {journal} {\bibinfo  {journal}
  {Phys. Rev. Lett.}\ }\textbf {\bibinfo {volume} {100}},\ \bibinfo {pages}
  {013605} (\bibinfo {year} {2008})}\BibitemShut {NoStop}%
\bibitem [{\citenamefont {Leggett}\ and\ \citenamefont
  {Garg}(1985)}]{leggett:1985}%
  \BibitemOpen
  \bibfield  {author} {\bibinfo {author} {\bibfnamefont {A.~J.}\ \bibnamefont
  {Leggett}}\ and\ \bibinfo {author} {\bibfnamefont {A.}~\bibnamefont {Garg}},\
  }\bibfield  {title} {\enquote {\bibinfo {title} {Quantum mechanics versus
  macroscopic realism: Is the flux there when nobody looks?}}\ }\href {\doibase
  10.1103/PhysRevLett.54.857} {\bibfield  {journal} {\bibinfo  {journal} {Phys.
  Rev. Lett.}\ }\textbf {\bibinfo {volume} {54}},\ \bibinfo {pages} {857}
  (\bibinfo {year} {1985})}\BibitemShut {NoStop}%
\bibitem [{\citenamefont {Emary}\ \emph {et~al.}(2014)\citenamefont {Emary},
  \citenamefont {Lambert},\ and\ \citenamefont {Nori}}]{emary:2014}%
  \BibitemOpen
  \bibfield  {author} {\bibinfo {author} {\bibfnamefont {C.}~\bibnamefont
  {Emary}}, \bibinfo {author} {\bibfnamefont {N.}~\bibnamefont {Lambert}}, \
  and\ \bibinfo {author} {\bibfnamefont {F.}~\bibnamefont {Nori}},\ }\bibfield
  {title} {\enquote {\bibinfo {title} {{L}eggett-{G}arg inequalities},}\ }\href
  {\doibase 10.1088/0034-4885/77/1/016001} {\bibfield  {journal} {\bibinfo
  {journal} {Rep. Prog. Phys.}\ }\textbf {\bibinfo {volume} {77}},\ \bibinfo
  {pages} {016001} (\bibinfo {year} {2014})}\BibitemShut {NoStop}%
\bibitem [{\citenamefont {von Neumann}(1932)}]{vonneumann:book}%
  \BibitemOpen
  \bibfield  {author} {\bibinfo {author} {\bibfnamefont {J.}~\bibnamefont {von
  Neumann}},\ }\href {\doibase 10.1007/978-3-642-61409-5} {\emph {\bibinfo
  {title} {Mathematische Grundlagen der Quantenmechanik}}}\ (\bibinfo
  {publisher} {Springer},\ \bibinfo {year} {1932})\BibitemShut {NoStop}%
\bibitem [{\citenamefont {Stenholm}(1992)}]{stenholm:1992}%
  \BibitemOpen
  \bibfield  {author} {\bibinfo {author} {\bibfnamefont {S.}~\bibnamefont
  {Stenholm}},\ }\bibfield  {title} {\enquote {\bibinfo {title} {Simultaneous
  measurement of conjugate variables},}\ }\href {\doibase
  10.1016/0003-4916(92)90086-2} {\bibfield  {journal} {\bibinfo  {journal}
  {Ann. Phys.}\ }\textbf {\bibinfo {volume} {218}},\ \bibinfo {pages} {233}
  (\bibinfo {year} {1992})}\BibitemShut {NoStop}%
\bibitem [{\citenamefont {Busch}(2009)}]{busch:2009}%
  \BibitemOpen
  \bibfield  {author} {\bibinfo {author} {\bibfnamefont {P.}~\bibnamefont
  {Busch}},\ }\enquote {\bibinfo {title} {`{N}o information without
  disturbance': Quantum limitations of measurement},}\ in\ \href {\doibase
  10.1007/978-1-4020-9107-0_13} {\emph {\bibinfo {booktitle} {Quantum Reality,
  Relativistic Causality, and Closing the Epistemic Circle: Essays in Honour of
  Abner Shimony}}}\ (\bibinfo  {publisher} {Springer Netherlands},\ \bibinfo
  {address} {Dordrecht},\ \bibinfo {year} {2009})\ p.\ \bibinfo {pages}
  {229}\BibitemShut {NoStop}%
\bibitem [{\citenamefont {Clerk}\ \emph
  {et~al.}(2010{\natexlab{b}})\citenamefont {Clerk}, \citenamefont
  {Marquardt},\ and\ \citenamefont {Harris}}]{clerk:2010}%
  \BibitemOpen
  \bibfield  {author} {\bibinfo {author} {\bibfnamefont {A.~A.}\ \bibnamefont
  {Clerk}}, \bibinfo {author} {\bibfnamefont {F.}~\bibnamefont {Marquardt}}, \
  and\ \bibinfo {author} {\bibfnamefont {J.~G.~E.}\ \bibnamefont {Harris}},\
  }\bibfield  {title} {\enquote {\bibinfo {title} {Quantum measurement of
  phonon shot noise},}\ }\href {\doibase 10.1103/PhysRevLett.104.213603}
  {\bibfield  {journal} {\bibinfo  {journal} {Phys. Rev. Lett.}\ }\textbf
  {\bibinfo {volume} {104}},\ \bibinfo {pages} {213603} (\bibinfo {year}
  {2010}{\natexlab{b}})}\BibitemShut {NoStop}%
\bibitem [{\citenamefont {Hofer}(2017)}]{hofer:2017}%
  \BibitemOpen
  \bibfield  {author} {\bibinfo {author} {\bibfnamefont {P.~P.}\ \bibnamefont
  {Hofer}},\ }\bibfield  {title} {\enquote {\bibinfo {title} {Quasi-probability
  distributions for observables in dynamic systems},}\ }\href {\doibase
  10.22331/q-2017-10-12-32} {\bibfield  {journal} {\bibinfo  {journal}
  {{Quantum}}\ }\textbf {\bibinfo {volume} {1}},\ \bibinfo {pages} {32}
  (\bibinfo {year} {2017})}\BibitemShut {NoStop}%
\bibitem [{\citenamefont {Aron}\ \emph {et~al.}(2018)\citenamefont {Aron},
  \citenamefont {Biroli},\ and\ \citenamefont {Cugliandolo}}]{aron:2018}%
  \BibitemOpen
  \bibfield  {author} {\bibinfo {author} {\bibfnamefont {C.}~\bibnamefont
  {Aron}}, \bibinfo {author} {\bibfnamefont {G.}~\bibnamefont {Biroli}}, \ and\
  \bibinfo {author} {\bibfnamefont {L.~F.}\ \bibnamefont {Cugliandolo}},\
  }\bibfield  {title} {\enquote {\bibinfo {title} {{(Non) equilibrium dynamics:
  a (broken) symmetry of the {K}eldysh generating functional}},}\ }\href
  {\doibase 10.21468/SciPostPhys.4.1.008} {\bibfield  {journal} {\bibinfo
  {journal} {SciPost Phys.}\ }\textbf {\bibinfo {volume} {4}},\ \bibinfo
  {pages} {8} (\bibinfo {year} {2018})}\BibitemShut {NoStop}%
\bibitem [{\citenamefont {Zachos}\ \emph {et~al.}(2005)\citenamefont {Zachos},
  \citenamefont {Fairlie},\ and\ \citenamefont {Curtright}}]{qm_phase:book}%
  \BibitemOpen
  \bibinfo {editor} {\bibfnamefont {C.~K.}\ \bibnamefont {Zachos}}, \bibinfo
  {editor} {\bibfnamefont {D.~B.}\ \bibnamefont {Fairlie}}, \ and\ \bibinfo
  {editor} {\bibfnamefont {T.~L.}\ \bibnamefont {Curtright}},\ eds.,\ \href
  {\doibase 10.1142/S0219749908003451} {\emph {\bibinfo {title} {Quantum
  Mechanics in Phase Space: An Overview with Selected Papers}}}\ (\bibinfo
  {publisher} {World Scientific},\ \bibinfo {year} {2005})\BibitemShut
  {NoStop}%
\bibitem [{\citenamefont {Nazarov}(2003)}]{nazarov:book2}%
  \BibitemOpen
  \bibinfo {editor} {\bibfnamefont {Yu.~V.}\ \bibnamefont {Nazarov}},\ ed.,\
  \href {\doibase 10.1007/978-94-010-0089-5} {\emph {\bibinfo {title} {Quantum
  Noise in Mesoscopic Physics}}}\ (\bibinfo  {publisher} {Springer},\ \bibinfo
  {year} {2003})\BibitemShut {NoStop}%
\bibitem [{\citenamefont {Esposito}\ \emph {et~al.}(2009)\citenamefont
  {Esposito}, \citenamefont {Harbola},\ and\ \citenamefont
  {Mukamel}}]{esposito:2009}%
  \BibitemOpen
  \bibfield  {author} {\bibinfo {author} {\bibfnamefont {M.}~\bibnamefont
  {Esposito}}, \bibinfo {author} {\bibfnamefont {U.}~\bibnamefont {Harbola}}, \
  and\ \bibinfo {author} {\bibfnamefont {S.}~\bibnamefont {Mukamel}},\
  }\bibfield  {title} {\enquote {\bibinfo {title} {Nonequilibrium fluctuations,
  fluctuation theorems, and counting statistics in quantum systems},}\ }\href
  {\doibase 10.1103/RevModPhys.81.1665} {\bibfield  {journal} {\bibinfo
  {journal} {Rev. Mod. Phys.}\ }\textbf {\bibinfo {volume} {81}},\ \bibinfo
  {pages} {1665} (\bibinfo {year} {2009})}\BibitemShut {NoStop}%
\bibitem [{\citenamefont {Solinas}\ and\ \citenamefont
  {Gasparinetti}(2015)}]{solinas:2015}%
  \BibitemOpen
  \bibfield  {author} {\bibinfo {author} {\bibfnamefont {P.}~\bibnamefont
  {Solinas}}\ and\ \bibinfo {author} {\bibfnamefont {S.}~\bibnamefont
  {Gasparinetti}},\ }\bibfield  {title} {\enquote {\bibinfo {title} {Full
  distribution of work done on a quantum system for arbitrary initial
  states},}\ }\href {\doibase 10.1103/PhysRevE.92.042150} {\bibfield  {journal}
  {\bibinfo  {journal} {Phys. Rev. E}\ }\textbf {\bibinfo {volume} {92}},\
  \bibinfo {pages} {042150} (\bibinfo {year} {2015})}\BibitemShut {NoStop}%
\bibitem [{\citenamefont {Solinas}\ and\ \citenamefont
  {Gasparinetti}(2016)}]{solinas:2016}%
  \BibitemOpen
  \bibfield  {author} {\bibinfo {author} {\bibfnamefont {P.}~\bibnamefont
  {Solinas}}\ and\ \bibinfo {author} {\bibfnamefont {S.}~\bibnamefont
  {Gasparinetti}},\ }\bibfield  {title} {\enquote {\bibinfo {title} {Probing
  quantum interference effects in the work distribution},}\ }\href {\doibase
  10.1103/PhysRevA.94.052103} {\bibfield  {journal} {\bibinfo  {journal} {Phys.
  Rev. A}\ }\textbf {\bibinfo {volume} {94}},\ \bibinfo {pages} {052103}
  (\bibinfo {year} {2016})}\BibitemShut {NoStop}%
\bibitem [{\citenamefont {B\"aumer}\ \emph {et~al.}()\citenamefont {B\"aumer},
  \citenamefont {Lostaglio}, \citenamefont {Perarnau-Llobet},\ and\
  \citenamefont {Sampaio}}]{baumer:2018}%
  \BibitemOpen
  \bibfield  {author} {\bibinfo {author} {\bibfnamefont {E.}~\bibnamefont
  {B\"aumer}}, \bibinfo {author} {\bibfnamefont {M.}~\bibnamefont {Lostaglio}},
  \bibinfo {author} {\bibfnamefont {M.}~\bibnamefont {Perarnau-Llobet}}, \ and\
  \bibinfo {author} {\bibfnamefont {R.}~\bibnamefont {Sampaio}},\ }\href@noop
  {} {\enquote {\bibinfo {title} {Fluctuating work in coherent quantum systems:
  proposals and limitations},}\ }\Eprint {http://arxiv.org/abs/1805.10096}
  {arXiv:1805.10096 [quant-ph]} \BibitemShut {NoStop}%
\bibitem [{\citenamefont {Chiara}\ \emph {et~al.}()\citenamefont {Chiara},
  \citenamefont {Solinas}, \citenamefont {Cerisola},\ and\ \citenamefont
  {Roncaglia}}]{chiara:2018}%
  \BibitemOpen
  \bibfield  {author} {\bibinfo {author} {\bibfnamefont {G.~De}\ \bibnamefont
  {Chiara}}, \bibinfo {author} {\bibfnamefont {P.}~\bibnamefont {Solinas}},
  \bibinfo {author} {\bibfnamefont {F.}~\bibnamefont {Cerisola}}, \ and\
  \bibinfo {author} {\bibfnamefont {A.~J.}\ \bibnamefont {Roncaglia}},\
  }\href@noop {} {\enquote {\bibinfo {title} {Ancilla-assisted measurement of
  quantum work},}\ }\Eprint {http://arxiv.org/abs/1805.06047} {arXiv:1805.06047
  [quant-ph]} \BibitemShut {NoStop}%
\bibitem [{\citenamefont {Lostaglio}(2018)}]{lostaglio:2018}%
  \BibitemOpen
  \bibfield  {author} {\bibinfo {author} {\bibfnamefont {M.}~\bibnamefont
  {Lostaglio}},\ }\bibfield  {title} {\enquote {\bibinfo {title} {Quantum
  fluctuation theorems, contextuality, and work quasiprobabilities},}\ }\href
  {\doibase 10.1103/PhysRevLett.120.040602} {\bibfield  {journal} {\bibinfo
  {journal} {Phys. Rev. Lett.}\ }\textbf {\bibinfo {volume} {120}},\ \bibinfo
  {pages} {040602} (\bibinfo {year} {2018})}\BibitemShut {NoStop}%
\bibitem [{\citenamefont {Clerk}(2011)}]{clerk:2011}%
  \BibitemOpen
  \bibfield  {author} {\bibinfo {author} {\bibfnamefont {A.~A.}\ \bibnamefont
  {Clerk}},\ }\bibfield  {title} {\enquote {\bibinfo {title} {Full counting
  statistics of energy fluctuations in a driven quantum resonator},}\ }\href
  {\doibase 10.1103/PhysRevA.84.043824} {\bibfield  {journal} {\bibinfo
  {journal} {Phys. Rev. A}\ }\textbf {\bibinfo {volume} {84}},\ \bibinfo
  {pages} {043824} (\bibinfo {year} {2011})}\BibitemShut {NoStop}%
\bibitem [{\citenamefont {Schwonnek}\ and\ \citenamefont
  {Werner}()}]{schwonnek:2018}%
  \BibitemOpen
  \bibfield  {author} {\bibinfo {author} {\bibfnamefont {R.}~\bibnamefont
  {Schwonnek}}\ and\ \bibinfo {author} {\bibfnamefont {R.~F.}\ \bibnamefont
  {Werner}},\ }\href@noop {} {\enquote {\bibinfo {title} {Wigner distributions
  for $n$ arbitrary operators},}\ }\Eprint {http://arxiv.org/abs/1802.08342}
  {arXiv:1802.08342 [quant-ph]} \BibitemShut {NoStop}%
\bibitem [{\citenamefont {Hallaji}\ \emph {et~al.}(2017)\citenamefont
  {Hallaji}, \citenamefont {Feizpour}, \citenamefont {Dmochowski},
  \citenamefont {Sinclair},\ and\ \citenamefont {Steinberg}}]{hallaji:2017}%
  \BibitemOpen
  \bibfield  {author} {\bibinfo {author} {\bibfnamefont {M.}~\bibnamefont
  {Hallaji}}, \bibinfo {author} {\bibfnamefont {A.}~\bibnamefont {Feizpour}},
  \bibinfo {author} {\bibfnamefont {G.}~\bibnamefont {Dmochowski}}, \bibinfo
  {author} {\bibfnamefont {J.}~\bibnamefont {Sinclair}}, \ and\ \bibinfo
  {author} {\bibfnamefont {A. M.}\ \bibnamefont {Steinberg}},\ }\bibfield
  {title} {\enquote {\bibinfo {title} {Weak-value amplification of the
  nonlinear effect of a single photon},}\ }\href
  {https://doi.org/10.1038/nphys4040} {\bibfield  {journal} {\bibinfo
  {journal} {Nat. Phys.}\ }\textbf {\bibinfo {volume} {13}},\ \bibinfo {pages}
  {540} (\bibinfo {year} {2017})}\BibitemShut {NoStop}%
\bibitem [{\citenamefont {Sinclair}\ \emph {et~al.}()\citenamefont {Sinclair},
  \citenamefont {Spierings}, \citenamefont {Brodutch},\ and\ \citenamefont
  {Steinberg}}]{sinclair:2018}%
  \BibitemOpen
  \bibfield  {author} {\bibinfo {author} {\bibfnamefont {J.}~\bibnamefont
  {Sinclair}}, \bibinfo {author} {\bibfnamefont {D.}~\bibnamefont {Spierings}},
  \bibinfo {author} {\bibfnamefont {A.}~\bibnamefont {Brodutch}}, \ and\
  \bibinfo {author} {\bibfnamefont {A.~M.}\ \bibnamefont {Steinberg}},\
  }\href@noop {} {\enquote {\bibinfo {title} {Weak values and neoclassical
  realism},}\ }\Eprint {http://arxiv.org/abs/1808.09951} {arXiv:1808.09951
  [quant-ph]} \BibitemShut {NoStop}%
\bibitem [{\citenamefont {Yunger~Halpern}\ \emph {et~al.}(2018)\citenamefont
  {Yunger~Halpern}, \citenamefont {Swingle},\ and\ \citenamefont
  {Dressel}}]{halpern:2018}%
  \BibitemOpen
  \bibfield  {author} {\bibinfo {author} {\bibfnamefont {N.}~\bibnamefont
  {Yunger~Halpern}}, \bibinfo {author} {\bibfnamefont {B.}~\bibnamefont
  {Swingle}}, \ and\ \bibinfo {author} {\bibfnamefont {J.}~\bibnamefont
  {Dressel}},\ }\bibfield  {title} {\enquote {\bibinfo {title}
  {Quasiprobability behind the out-of-time-ordered correlator},}\ }\href
  {\doibase 10.1103/PhysRevA.97.042105} {\bibfield  {journal} {\bibinfo
  {journal} {Phys. Rev. A}\ }\textbf {\bibinfo {volume} {97}},\ \bibinfo
  {pages} {042105} (\bibinfo {year} {2018})}\BibitemShut {NoStop}%
\bibitem [{\citenamefont {Belzig}\ and\ \citenamefont
  {Nazarov}(2001)}]{belzig:2001}%
  \BibitemOpen
  \bibfield  {author} {\bibinfo {author} {\bibfnamefont {W.}~\bibnamefont
  {Belzig}}\ and\ \bibinfo {author} {\bibfnamefont {Yu.~V.}\ \bibnamefont
  {Nazarov}},\ }\bibfield  {title} {\enquote {\bibinfo {title} {Full counting
  statistics of electron transfer between superconductors},}\ }\href {\doibase
  10.1103/PhysRevLett.87.197006} {\bibfield  {journal} {\bibinfo  {journal}
  {Phys. Rev. Lett.}\ }\textbf {\bibinfo {volume} {87}},\ \bibinfo {pages}
  {197006} (\bibinfo {year} {2001})}\BibitemShut {NoStop}%
\bibitem [{\citenamefont {Kenfack}\ and\ \citenamefont
  {{\.Z}yczkowski}(2004)}]{kenfack:2004}%
  \BibitemOpen
  \bibfield  {author} {\bibinfo {author} {\bibfnamefont {A.}~\bibnamefont
  {Kenfack}}\ and\ \bibinfo {author} {\bibfnamefont {K.}~\bibnamefont
  {{\.Z}yczkowski}},\ }\bibfield  {title} {\enquote {\bibinfo {title}
  {Negativity of the {W}igner function as an indicator of non-classicality},}\
  }\href {\doibase 10.1088/1464-4266/6/10/003} {\bibfield  {journal} {\bibinfo
  {journal} {J. Opt. B}\ }\textbf {\bibinfo {volume} {6}},\ \bibinfo {pages}
  {396} (\bibinfo {year} {2004})}\BibitemShut {NoStop}%
\bibitem [{\citenamefont {Deleglise}\ \emph {et~al.}(2008)\citenamefont
  {Deleglise}, \citenamefont {Dotsenko}, \citenamefont {Sayrin}, \citenamefont
  {Bernu}, \citenamefont {Brune}, \citenamefont {Raimond},\ and\ \citenamefont
  {Haroche}}]{deleglise:2008}%
  \BibitemOpen
  \bibfield  {author} {\bibinfo {author} {\bibfnamefont {S.}~\bibnamefont
  {Deleglise}}, \bibinfo {author} {\bibfnamefont {I.}~\bibnamefont {Dotsenko}},
  \bibinfo {author} {\bibfnamefont {C.}~\bibnamefont {Sayrin}}, \bibinfo
  {author} {\bibfnamefont {J.}~\bibnamefont {Bernu}}, \bibinfo {author}
  {\bibfnamefont {M.}~\bibnamefont {Brune}}, \bibinfo {author} {\bibfnamefont
  {J.-M.}\ \bibnamefont {Raimond}}, \ and\ \bibinfo {author} {\bibfnamefont
  {S.}~\bibnamefont {Haroche}},\ }\bibfield  {title} {\enquote {\bibinfo
  {title} {Reconstruction of non-classical cavity field states with snapshots
  of their decoherence},}\ }\href {\doibase 10.1038/nature07288} {\bibfield
  {journal} {\bibinfo  {journal} {Nature}\ }\textbf {\bibinfo {volume} {455}},\
  \bibinfo {pages} {510} (\bibinfo {year} {2008})}\BibitemShut {NoStop}%
\bibitem [{\citenamefont {Bednorz}\ and\ \citenamefont
  {Belzig}(2010)}]{bednorz:2010}%
  \BibitemOpen
  \bibfield  {author} {\bibinfo {author} {\bibfnamefont {A.}~\bibnamefont
  {Bednorz}}\ and\ \bibinfo {author} {\bibfnamefont {W.}~\bibnamefont
  {Belzig}},\ }\bibfield  {title} {\enquote {\bibinfo {title}
  {Quasiprobabilistic interpretation of weak measurements in mesoscopic
  junctions},}\ }\href {\doibase 10.1103/PhysRevLett.105.106803} {\bibfield
  {journal} {\bibinfo  {journal} {Phys. Rev. Lett.}\ }\textbf {\bibinfo
  {volume} {105}},\ \bibinfo {pages} {106803} (\bibinfo {year}
  {2010})}\BibitemShut {NoStop}%
\bibitem [{\citenamefont {Bednorz}\ \emph {et~al.}(2012)\citenamefont
  {Bednorz}, \citenamefont {Belzig},\ and\ \citenamefont
  {Nitzan}}]{bednorz:2012}%
  \BibitemOpen
  \bibfield  {author} {\bibinfo {author} {\bibfnamefont {A.}~\bibnamefont
  {Bednorz}}, \bibinfo {author} {\bibfnamefont {W.}~\bibnamefont {Belzig}}, \
  and\ \bibinfo {author} {\bibfnamefont {A.}~\bibnamefont {Nitzan}},\
  }\bibfield  {title} {\enquote {\bibinfo {title} {Nonclassical time
  correlation functions in continuous quantum measurement},}\ }\href {\doibase
  10.1088/1367-2630/14/1/013009} {\bibfield  {journal} {\bibinfo  {journal}
  {New J. Phys.}\ }\textbf {\bibinfo {volume} {14}},\ \bibinfo {pages} {013009}
  (\bibinfo {year} {2012})}\BibitemShut {NoStop}%
\bibitem [{\citenamefont {Hofer}\ and\ \citenamefont
  {Clerk}(2016)}]{hofer:2016}%
  \BibitemOpen
  \bibfield  {author} {\bibinfo {author} {\bibfnamefont {P.~P.}\ \bibnamefont
  {Hofer}}\ and\ \bibinfo {author} {\bibfnamefont {A.~A.}\ \bibnamefont
  {Clerk}},\ }\bibfield  {title} {\enquote {\bibinfo {title} {Negative full
  counting statistics arise from interference effects},}\ }\href {\doibase
  10.1103/PhysRevLett.116.013603} {\bibfield  {journal} {\bibinfo  {journal}
  {Phys. Rev. Lett.}\ }\textbf {\bibinfo {volume} {116}},\ \bibinfo {pages}
  {013603} (\bibinfo {year} {2016})}\BibitemShut {NoStop}%
\bibitem [{\citenamefont {Kent}(1999)}]{kent:1999}%
  \BibitemOpen
  \bibfield  {author} {\bibinfo {author} {\bibfnamefont {A.}~\bibnamefont
  {Kent}},\ }\bibfield  {title} {\enquote {\bibinfo {title} {Noncontextual
  hidden variables and physical measurements},}\ }\href {\doibase
  10.1103/PhysRevLett.83.3755} {\bibfield  {journal} {\bibinfo  {journal}
  {Phys. Rev. Lett.}\ }\textbf {\bibinfo {volume} {83}},\ \bibinfo {pages}
  {3755} (\bibinfo {year} {1999})}\BibitemShut {NoStop}%
\bibitem [{\citenamefont {Meyer}(1999)}]{meyer:1999}%
  \BibitemOpen
  \bibfield  {author} {\bibinfo {author} {\bibfnamefont {David~A.}\
  \bibnamefont {Meyer}},\ }\bibfield  {title} {\enquote {\bibinfo {title}
  {Finite precision measurement nullifies the {K}ochen-{S}pecker theorem},}\
  }\href {\doibase 10.1103/PhysRevLett.83.3751} {\bibfield  {journal} {\bibinfo
   {journal} {Phys. Rev. Lett.}\ }\textbf {\bibinfo {volume} {83}},\ \bibinfo
  {pages} {3751} (\bibinfo {year} {1999})}\BibitemShut {NoStop}%
\bibitem [{\citenamefont {Spekkens}(2008)}]{spekkens:2008}%
  \BibitemOpen
  \bibfield  {author} {\bibinfo {author} {\bibfnamefont {Robert~W.}\
  \bibnamefont {Spekkens}},\ }\bibfield  {title} {\enquote {\bibinfo {title}
  {Negativity and contextuality are equivalent notions of nonclassicality},}\
  }\href {\doibase 10.1103/PhysRevLett.101.020401} {\bibfield  {journal}
  {\bibinfo  {journal} {Phys. Rev. Lett.}\ }\textbf {\bibinfo {volume} {101}},\
  \bibinfo {pages} {020401} (\bibinfo {year} {2008})}\BibitemShut {NoStop}%
\bibitem [{\citenamefont {Cabello}(2008)}]{cabello:2008}%
  \BibitemOpen
  \bibfield  {author} {\bibinfo {author} {\bibfnamefont {A.}~\bibnamefont
  {Cabello}},\ }\bibfield  {title} {\enquote {\bibinfo {title} {Experimentally
  testable state-independent quantum contextuality},}\ }\href {\doibase
  10.1103/PhysRevLett.101.210401} {\bibfield  {journal} {\bibinfo  {journal}
  {Phys. Rev. Lett.}\ }\textbf {\bibinfo {volume} {101}},\ \bibinfo {pages}
  {210401} (\bibinfo {year} {2008})}\BibitemShut {NoStop}%
\bibitem [{\citenamefont {Kirchmair}\ \emph {et~al.}(2009)\citenamefont
  {Kirchmair}, \citenamefont {Z\"ahringer}, \citenamefont {Gerritsma},
  \citenamefont {Kleinmann}, \citenamefont {G\"uhne}, \citenamefont {Cabello},
  \citenamefont {Blatt},\ and\ \citenamefont {Roos}}]{kirchmair:2009}%
  \BibitemOpen
  \bibfield  {author} {\bibinfo {author} {\bibfnamefont {G.}~\bibnamefont
  {Kirchmair}}, \bibinfo {author} {\bibfnamefont {F.}~\bibnamefont
  {Z\"ahringer}}, \bibinfo {author} {\bibfnamefont {R.}~\bibnamefont
  {Gerritsma}}, \bibinfo {author} {\bibfnamefont {M.}~\bibnamefont
  {Kleinmann}}, \bibinfo {author} {\bibfnamefont {O.}~\bibnamefont {G\"uhne}},
  \bibinfo {author} {\bibfnamefont {A.}~\bibnamefont {Cabello}}, \bibinfo
  {author} {\bibfnamefont {R.}~\bibnamefont {Blatt}}, \ and\ \bibinfo {author}
  {\bibfnamefont {C.~F.}\ \bibnamefont {Roos}},\ }\bibfield  {title} {\enquote
  {\bibinfo {title} {State-independent experimental test of quantum
  contextuality},}\ }\href {https://doi.org/10.1038/nature08172} {\bibfield
  {journal} {\bibinfo  {journal} {Nature}\ }\textbf {\bibinfo {volume} {460}},\
  \bibinfo {pages} {494} (\bibinfo {year} {2009})}\BibitemShut {NoStop}%
\bibitem [{\citenamefont {B\"ulte}\ \emph {et~al.}(2018)\citenamefont
  {B\"ulte}, \citenamefont {Bednorz}, \citenamefont {Bruder},\ and\
  \citenamefont {Belzig}}]{bulte:2018}%
  \BibitemOpen
  \bibfield  {author} {\bibinfo {author} {\bibfnamefont {J.}~\bibnamefont
  {B\"ulte}}, \bibinfo {author} {\bibfnamefont {A.}~\bibnamefont {Bednorz}},
  \bibinfo {author} {\bibfnamefont {C.}~\bibnamefont {Bruder}}, \ and\ \bibinfo
  {author} {\bibfnamefont {W.}~\bibnamefont {Belzig}},\ }\bibfield  {title}
  {\enquote {\bibinfo {title} {Noninvasive quantum measurement of arbitrary
  operator order by engineered non-{M}arkovian detectors},}\ }\href {\doibase
  10.1103/PhysRevLett.120.140407} {\bibfield  {journal} {\bibinfo  {journal}
  {Phys. Rev. Lett.}\ }\textbf {\bibinfo {volume} {120}},\ \bibinfo {pages}
  {140407} (\bibinfo {year} {2018})}\BibitemShut {NoStop}%
\bibitem [{\citenamefont {Nazarov}\ and\ \citenamefont
  {Kindermann}(2003)}]{nazarov:2003}%
  \BibitemOpen
  \bibfield  {author} {\bibinfo {author} {\bibfnamefont {{Yu}.~V.}\
  \bibnamefont {Nazarov}}\ and\ \bibinfo {author} {\bibfnamefont
  {M.}~\bibnamefont {Kindermann}},\ }\bibfield  {title} {\enquote {\bibinfo
  {title} {Full counting statistics of a general quantum mechanical
  variable},}\ }\href {\doibase 10.1140/epjb/e2003-00293-1} {\bibfield
  {journal} {\bibinfo  {journal} {Eur. Phys. J. B}\ }\textbf {\bibinfo {volume}
  {35}},\ \bibinfo {pages} {413} (\bibinfo {year} {2003})}\BibitemShut
  {NoStop}%
\bibitem [{\citenamefont {Di~Lorenzo}(2011)}]{di_lorenzo:2011}%
  \BibitemOpen
  \bibfield  {author} {\bibinfo {author} {\bibfnamefont {A.}~\bibnamefont
  {Di~Lorenzo}},\ }\bibfield  {title} {\enquote {\bibinfo {title} {Strong
  correspondence principle for joint measurement of conjugate observables},}\
  }\href {\doibase 10.1103/PhysRevA.83.042104} {\bibfield  {journal} {\bibinfo
  {journal} {Phys. Rev. A}\ }\textbf {\bibinfo {volume} {83}},\ \bibinfo
  {pages} {042104} (\bibinfo {year} {2011})}\BibitemShut {NoStop}%
\bibitem [{\citenamefont {Barnea}\ \emph {et~al.}(2017)\citenamefont {Barnea},
  \citenamefont {Renou}, \citenamefont {Fr\"owis},\ and\ \citenamefont
  {Gisin}}]{barnea:2017}%
  \BibitemOpen
  \bibfield  {author} {\bibinfo {author} {\bibfnamefont {T.~J.}\ \bibnamefont
  {Barnea}}, \bibinfo {author} {\bibfnamefont {M.-O.}\ \bibnamefont {Renou}},
  \bibinfo {author} {\bibfnamefont {F.}~\bibnamefont {Fr\"owis}}, \ and\
  \bibinfo {author} {\bibfnamefont {N.}~\bibnamefont {Gisin}},\ }\bibfield
  {title} {\enquote {\bibinfo {title} {Macroscopic quantum measurements of
  noncommuting observables},}\ }\href {\doibase 10.1103/PhysRevA.96.012111}
  {\bibfield  {journal} {\bibinfo  {journal} {Phys. Rev. A}\ }\textbf {\bibinfo
  {volume} {96}},\ \bibinfo {pages} {012111} (\bibinfo {year}
  {2017})}\BibitemShut {NoStop}%
\bibitem [{sup()}]{supp}%
  \BibitemOpen
  \href@noop {} {}\bibinfo {note} {See supplemental information for a detailed
  derivation of the proposed inequality and detailed calculations for the
  examples in the main text.}\BibitemShut {Stop}%
\bibitem [{\citenamefont {Arthurs}\ and\ \citenamefont
  {Kelly}(1965)}]{arthurs:1964}%
  \BibitemOpen
  \bibfield  {author} {\bibinfo {author} {\bibfnamefont {E.}~\bibnamefont
  {Arthurs}}\ and\ \bibinfo {author} {\bibfnamefont {J.~L.}\ \bibnamefont
  {Kelly}},\ }\bibfield  {title} {\enquote {\bibinfo {title} {B.{S}.{T}.{J}.
  briefs: On the simultaneous measurement of a pair of conjugate
  observables},}\ }\href {\doibase 10.1002/j.1538-7305.1965.tb01684.x}
  {\bibfield  {journal} {\bibinfo  {journal} {Bell Syst. Tech. J.}\ }\textbf
  {\bibinfo {volume} {44}},\ \bibinfo {pages} {725} (\bibinfo {year}
  {1965})}\BibitemShut {NoStop}%
\bibitem [{\citenamefont {Leonhardt}\ and\ \citenamefont
  {Paul}(1995)}]{leonhardt:1995}%
  \BibitemOpen
  \bibfield  {author} {\bibinfo {author} {\bibfnamefont {U.}~\bibnamefont
  {Leonhardt}}\ and\ \bibinfo {author} {\bibfnamefont {H.}~\bibnamefont
  {Paul}},\ }\bibfield  {title} {\enquote {\bibinfo {title} {Measuring the
  quantum state of light},}\ }\href {\doibase 10.1016/0079-6727(94)00007-L}
  {\bibfield  {journal} {\bibinfo  {journal} {Prog. Quant. Electron.}\ }\textbf
  {\bibinfo {volume} {19}},\ \bibinfo {pages} {89} (\bibinfo {year}
  {1995})}\BibitemShut {NoStop}%
\bibitem [{\citenamefont {Eichler}\ \emph {et~al.}(2011)\citenamefont
  {Eichler}, \citenamefont {Bozyigit}, \citenamefont {Lang}, \citenamefont
  {Steffen}, \citenamefont {Fink},\ and\ \citenamefont
  {Wallraff}}]{eichler:2011}%
  \BibitemOpen
  \bibfield  {author} {\bibinfo {author} {\bibfnamefont {C.}~\bibnamefont
  {Eichler}}, \bibinfo {author} {\bibfnamefont {D.}~\bibnamefont {Bozyigit}},
  \bibinfo {author} {\bibfnamefont {C.}~\bibnamefont {Lang}}, \bibinfo {author}
  {\bibfnamefont {L.}~\bibnamefont {Steffen}}, \bibinfo {author} {\bibfnamefont
  {J.}~\bibnamefont {Fink}}, \ and\ \bibinfo {author} {\bibfnamefont
  {A.}~\bibnamefont {Wallraff}},\ }\bibfield  {title} {\enquote {\bibinfo
  {title} {Experimental state tomography of itinerant single microwave
  photons},}\ }\href {\doibase 10.1103/PhysRevLett.106.220503} {\bibfield
  {journal} {\bibinfo  {journal} {Phys. Rev. Lett.}\ }\textbf {\bibinfo
  {volume} {106}},\ \bibinfo {pages} {220503} (\bibinfo {year}
  {2011})}\BibitemShut {NoStop}%
\bibitem [{\citenamefont {Feuerverger}\ and\ \citenamefont
  {Mureika}(1977)}]{feuerverger:1977}%
  \BibitemOpen
  \bibfield  {author} {\bibinfo {author} {\bibfnamefont {A.}~\bibnamefont
  {Feuerverger}}\ and\ \bibinfo {author} {\bibfnamefont {R.~A.}\ \bibnamefont
  {Mureika}},\ }\bibfield  {title} {\enquote {\bibinfo {title} {The empirical
  characteristic function and its applications},}\ }\href {\doibase
  doi:10.1214/aos/1176343742} {\bibfield  {journal} {\bibinfo  {journal} {Ann.
  Statist.}\ }\textbf {\bibinfo {volume} {5}},\ \bibinfo {pages} {88} (\bibinfo
  {year} {1977})}\BibitemShut {NoStop}%
\bibitem [{\citenamefont {Hacohen-Gourgy}\ \emph {et~al.}(2016)\citenamefont
  {Hacohen-Gourgy}, \citenamefont {Martin}, \citenamefont {Flurin},
  \citenamefont {Ramasesh}, \citenamefont {Whaley},\ and\ \citenamefont
  {Siddiqi}}]{hacohen:2016}%
  \BibitemOpen
  \bibfield  {author} {\bibinfo {author} {\bibfnamefont {S.}~\bibnamefont
  {Hacohen-Gourgy}}, \bibinfo {author} {\bibfnamefont {L.~S.}\ \bibnamefont
  {Martin}}, \bibinfo {author} {\bibfnamefont {E.}~\bibnamefont {Flurin}},
  \bibinfo {author} {\bibfnamefont {V.~V.}\ \bibnamefont {Ramasesh}}, \bibinfo
  {author} {\bibfnamefont {K.~B.}\ \bibnamefont {Whaley}}, \ and\ \bibinfo
  {author} {\bibfnamefont {I.}~\bibnamefont {Siddiqi}},\ }\bibfield  {title}
  {\enquote {\bibinfo {title} {Quantum dynamics of simultaneously measured
  non-commuting observables},}\ }\href {\doibase 10.1038/nature19762}
  {\bibfield  {journal} {\bibinfo  {journal} {Nature}\ }\textbf {\bibinfo
  {volume} {538}},\ \bibinfo {pages} {491} (\bibinfo {year}
  {2016})}\BibitemShut {NoStop}%
\bibitem [{\citenamefont {Perarnau-Llobet}\ and\ \citenamefont
  {Nieuwenhuizen}(2017)}]{perarnau:2017}%
  \BibitemOpen
  \bibfield  {author} {\bibinfo {author} {\bibfnamefont {M.}~\bibnamefont
  {Perarnau-Llobet}}\ and\ \bibinfo {author} {\bibfnamefont {T.~M.}\
  \bibnamefont {Nieuwenhuizen}},\ }\bibfield  {title} {\enquote {\bibinfo
  {title} {Simultaneous measurement of two noncommuting quantum variables:
  Solution of a dynamical model},}\ }\href {\doibase
  10.1103/PhysRevA.95.052129} {\bibfield  {journal} {\bibinfo  {journal} {Phys.
  Rev. A}\ }\textbf {\bibinfo {volume} {95}},\ \bibinfo {pages} {052129}
  (\bibinfo {year} {2017})}\BibitemShut {NoStop}%
\end{thebibliography}%

\clearpage
\widetext

\widetext
\begin{center}
	\textbf{\large Supplemental information: Certifying Non-Classical Behavior for Negative Keldysh Quasi-Probabilities}
\end{center}
\setcounter{equation}{0}
\setcounter{figure}{0}
\setcounter{table}{0}
\setcounter{page}{1}
\makeatletter
\renewcommand{\theequation}{S\arabic{equation}}
\renewcommand{\thefigure}{S\arabic{figure}}
Here we provide supplementary calculations and expressions for the examples discussed in the main text. Equation and Figure numbers not preceded by an `$S$' refer to the main text.

\section{A. Detailed derivation of the inequality}

Here we give a detailed derivation of the inequality $K_{\rm cl}\geq0$ [Eq.~\eqref{eq:inequality} in the main text], illustrating where the assumptions on the detectors enter the derivation. Our starting point is given by Eq.~\eqref{eq:reckqpd} in the main text which reads
\begin{equation}
\label{eq:appkcl}
K_{\rm cl} = \frac{1}{(2\pi)^2}\int d\boldsymbol{\lambda}e^{i\boldsymbol{\lambda}\cdot\boldsymbol{A}}\tilde{P}_{\rm cl}(\boldsymbol{\lambda}|\boldsymbol{\chi})\prod_{j=1,2}\frac{\tilde{P}_{\rm cl}(\lambda_j|\chi_j')}{\tilde{P}_{\rm cl}(\lambda_j|\chi_j)},
\end{equation}
where 
\begin{equation}
\label{eq:plam}
\tilde{P}_{\rm cl}(\boldsymbol{\lambda}|\boldsymbol{\chi})=\int d\boldsymbol{A}e^{-i\boldsymbol{\lambda}\cdot\boldsymbol{A}}P_{\rm cl}(\boldsymbol{A}|\boldsymbol{\chi})\hspace{1cm}\tilde{P}_{\rm cl}(\lambda_j|\chi_j)=\int d A_je^{-i\lambda_jA_j}P_{\rm cl}(A_j|\chi_j).
\end{equation}
The distribution describing the joint measurement of $A_1$ and $A_2$ is given by
\begin{equation}
\label{eq:appclassgen}
P_{\rm cl}(\boldsymbol{A}|\boldsymbol{\chi}) = \int d\boldsymbol{A}'d\boldsymbol{\gamma} \left[\prod_j M_j({A}_j,{A}'_j,\gamma_j|\chi_j)\right]S(\boldsymbol{A}'|\boldsymbol{\gamma}).
\end{equation}
Here we have already made the assumption of uncorrelated detectors. While this assumption is necessary only later in the derivation, the whole derivation becomes considerably more transparent by making the assumption at this early stage. The quantity $P_{\rm cl}(A_j|\chi_j)$ describes the measurement of a single observable and is not in general given by the marginal of Eq.~\eqref{eq:appclassgen}. To recover $P_{\rm cl}(A_j|\chi_j)$ from $P_{\rm cl}(\boldsymbol{A}|\boldsymbol{\chi})$, we require assumption five (detectors can be detached)
\begin{equation}
\lim_{\chi_j\rightarrow 0}M_j({A}_j,{A}'_j,\gamma_j|\chi_j)=\delta(\gamma_j)U(A_j).
\end{equation}
With this equation, we find
\begin{equation}
\label{eq:singleobs}
P_{\rm cl}(A_j|\chi_j)=\lim_{\chi_k\rightarrow 0}\int dA_kP_{\rm cl}(\boldsymbol{A}|\boldsymbol{\chi}) = \int d{A}'_jd\gamma_j M_j({A}_j,{A}'_j,\gamma_j|\chi_j)S(A'_j|\gamma_j),
\end{equation}
where $k\neq j$ and, as in the main text, we defined $\int dA_{k}S(\boldsymbol{A}|\gamma_j,\gamma_k=0)\equiv S({A}_j|\gamma_j)$. Using the translational invariance of the detectors $M_j({A}_j,{A}'_j,\gamma_j|\chi_j)=M_j({A}_j-{A}'_j,\gamma_j|\chi_j)$, i.e., assumption 4, the Fourier transforms of the probability distributions reduce to products due to the convolution theorem
\begin{equation}
\label{eq:fourierprod}
\tilde{P}_{\rm cl}(\boldsymbol{\lambda}|\boldsymbol{\chi})=\int d\boldsymbol{\gamma}\left[\prod_j \tilde{M}_j(\lambda_j,\gamma_j|\chi_j)\right]\tilde{S}(\boldsymbol{\lambda}|\boldsymbol{\gamma}), \hspace{1cm}\tilde{P}_{\rm cl}(\lambda_j|\chi_j)=\int d\gamma_j\tilde{M}_j(\lambda_j,\gamma_j|\chi_j)\tilde{S}(\lambda_j|\gamma_j),
\end{equation}
where the Fourier transforms of $M$ and $S$ are given analogously to Eq.~\eqref{eq:plam}. Using Assumption 3 [a measurement of a single observable is not affected by backaction, i.e., $\tilde{S}(\lambda_j|\gamma_j)$ is independent of $\gamma_j$] we can write
\begin{equation}
\tilde{P}_{\rm cl}(\lambda_j|\chi_j)=\int d\gamma_j\tilde{M}_j(\lambda_j,\gamma_j|\chi_j)\tilde{S}(\lambda_j|\gamma_j)=\tilde{D}_j(\lambda_j|\chi_j)\tilde{S}(\lambda_j),
\end{equation}
where at this point, $D_j$ is obtained by integrating $M_j$ over $\gamma_j$. We can then write
\begin{equation}
\label{eq:appkcl2}
K_{\rm cl} = \frac{1}{(2\pi)^2}\int d\boldsymbol{\lambda}e^{i\boldsymbol{\lambda}\cdot\boldsymbol{A}}\int d\boldsymbol{\gamma}\left[\prod_j \tilde{M}_j(\lambda_j,\gamma_j|\chi_j)\frac{\tilde{D}_j(\lambda_j|\chi'_j)}{\tilde{D}_j(\lambda_j|\chi_j)}\right]\tilde{S}(\boldsymbol{\lambda}|\boldsymbol{\gamma}).
\end{equation}
Using assumption 2 (uncorrelated imprecision and backaction)
\begin{equation}
\tilde{M}_j(\lambda_j,\gamma_j|\chi_j)=p_j(\gamma_j|\chi_j)\tilde{D}_j(\lambda_j|\chi_j),
\end{equation}
Eq.~\eqref{eq:appkcl2} reduces to (this is where we also need assumption 1, uncorrelated detectors)
\begin{equation}
\label{eq:appkcl3}
\begin{aligned}
K_{\rm cl} = &\frac{1}{(2\pi)^2}\int d\boldsymbol{\lambda}e^{i\boldsymbol{\lambda}\cdot\boldsymbol{A}}\int d\boldsymbol{\gamma}\left[\prod_jp_j(\gamma_j|\chi_j) \tilde{D}_j(\lambda_j|\chi'_j)\right]\tilde{S}(\boldsymbol{\lambda}|\boldsymbol{\gamma})\\=&\int d\boldsymbol{A}'d\boldsymbol{\gamma}S(\boldsymbol{A}'|\boldsymbol{\gamma}) \prod_{j=1,2}p_j(\gamma_j|\chi_j)D_j({A}_j-{A}'_j|\chi_j')\geq 0.
\end{aligned}
\end{equation}
The second line corresponds to Eq.~\eqref{eq:reccl} in the main text and we used the convolution theorem once more to arrive there. Since all involved distributions are assumed to be positive, the inequality follows directly.

Note that within our assumptions, we can write
\begin{equation}
\tilde{P}_{\rm cl}(\boldsymbol{\lambda}|\boldsymbol{\chi})= \int d\boldsymbol{A}'d\boldsymbol{\gamma}S(\boldsymbol{A}'|\boldsymbol{\gamma}) \prod_{j=1,2}p_j(\gamma_j|\chi_j)D_j({A}_j-{A}'_j|\chi_j),
\end{equation}
which looks very similar to Eq.~\eqref{eq:appkcl3} with the sole exception that both the backaction ($p_j$) and the measurement imprecision ($D_j$) are determined by the same measurement strength ($\chi_j$). Intuitively, our assumptions allow for determining the measurement imprecision of a detector by measuring a single observable. In particular, the ratio
\begin{equation}
\frac{\tilde{P}_{\rm cl}(\lambda_j|\chi_j')}{\tilde{P}_{\rm cl}(\lambda_j|\chi_j)}=\frac{D_j(\lambda_j|\chi_j')}{D_j(\lambda_j|\chi_j)},
\end{equation}
is then just the ratios of the measurement imprecision terms. This allows for exchanging the measurement imprecision at one value of $\chi$ with the measurement imprecision at another value $\chi'$ which is how $K_{\rm cl}$ is related to $P_{\rm cl}$.

In the quantum case, $K$ can be obtained analogously but starting with Eq.~\eqref{eq:measqm} in the main text instead of Eq.~\eqref{eq:appclassgen}. One then proceeds by making the same assumptions on the detectors which corresponds to using Wigner functions that factor in a position and a momentum part.

\section{B. Simultaneous position and momentum measurements}

From Eqs.~\eqref{eq:measqm}
, \eqref{eq:wignerdet}
, and \eqref{eq:wignerfock}
, we find the probability distribution describing the outcomes of a joint position and momentum measurement
\begin{equation}
\label{eq:probaxpsim}
P(x,p|\chi)=\frac{4\chi^2}{\pi(2+\chi^2)^6}e^{-\frac{4\chi^2(x^2+p^2)}{(2+\chi^2)^2}}\left[32\chi^4(x^2+p^2)+(4-\chi^4)^2\right],
\end{equation}
where we assumed the measurement strengths to be equal, i.e., $\chi_x=\chi_p=\chi$. The corresponding characteristic function reads
\begin{equation}
\label{eq:charaxpsim}
\tilde{P}(\lambda_x,\lambda_p|\chi)=\int dxdpe^{-i\lambda_x x-i\lambda_p p}P(x,p|\chi)=\frac{1}{2}e^{-\frac{(2+\chi^2)^2}{16\chi^2}(\lambda_x^2+\lambda_p^2)}(2-\lambda_x^2-\lambda_p^2).
\end{equation}
Analogously, we find the probability distribution that describes a measurement of $\hat{x}$ alone
\begin{equation}
\label{eq:probax}
P(x|\chi)=\frac{\chi}{\sqrt{\pi(1+\chi^2)^5}}(1+\chi^2+2x^2\chi^4)e^{-\frac{\chi^2x^2}{1+\chi^2}},
\end{equation}
with the characteristic function
\begin{equation}
\label{eq:probaxchar}
\tilde{P}(\lambda_x|\chi)=\frac{1}{2}e^{-\frac{1+\chi^2}{4\chi^2}\lambda_x^2}(2-\lambda_x^2).
\end{equation} 
Due to the rotational invariance of the Wigner function of a single-photon Fock state [cf.~Eq.~\eqref{eq:wignerfock}], the distributions for a measurement of momentum alone are equivalent. From the last equation, we find
\begin{equation}
\label{eq:measimpxp}
\frac{\tilde{P}(\lambda_x|\chi)}{\tilde{P}(\lambda_x|\chi')}=e^{-\frac{\lambda^2}{4\chi^2}[1-(\chi/\chi')^2]}=\frac{\tilde{D}(\lambda_x|\chi)}{\tilde{D}(\lambda_x|\chi')},
\end{equation}
where for the last equality, we used the Wigner function of the detectors [cf.~Eq.~\eqref{eq:wignerdet}], and the fact that they can be written as
\begin{equation}
\label{eq:detimpwig}
\mathcal{W}_j(\chi_j{A}_j,{\gamma}_j/\chi_j) = \frac{1}{\sqrt{\pi}\chi_j}e^{-\gamma_j^2/\chi_j^2}\frac{\chi_j}{\sqrt{\pi}}e^{-\chi_j^2A_j^2}= p(\gamma_j|\chi_j)D(A_j|\chi_j),
\end{equation}
where, as discussed in the main text, $p(\gamma_j|\chi_j)$ encodes the backaction (arising from the momentum distribution of the detector) and $D(A_j|\chi_j)$ encodes the imprecision (arising from the position distribution). Equation \eqref{eq:measimpxp} thus shows that the measurement imprecision of a detector can be isolated by measuring a single observable with different measurement strengths. Importantly, this works because backaction is irrelevant when measuring a single observable (we are not interested in the post-measured state). We can now write
\begin{equation}
\label{eq:recxp}
K(x,p) = \int dx'dp'd\gamma_x d\gamma_p\mathcal{P}(x,p|\gamma_x,\gamma_p) \left[p(\gamma_x|\chi)D(x-x'|\chi')\right]\left[p(\gamma_p|\chi)D(p-p'|\chi')\right],
\end{equation}
where $\mathcal{P}(x,p|\gamma_x,\gamma_p)=\mathcal{W}(x-\gamma_p/2,p+\gamma_x/2)$. Note the close similarity between Eq.~\eqref{eq:reccl}
in the main text and Eq.~\eqref{eq:recxp}.
From Eqs.~\eqref{eq:charaxpsim} and \eqref{eq:measimpxp}, we recover $K$ given in Eq.~\eqref{eq:kfock}
in the main text.

\section{C. Subsequent measurements on a two-level system}

In the second example, we consider a two level system in a pure state
\begin{equation}
\label{eq:initial}
\hat{\rho}=|+\rangle,\hspace{2cm}\hat{\sigma}_x|+\rangle=|+\rangle,
\end{equation}
where $\hat{\sigma}_x$ denotes a Pauli matrix. We are then interested in a weak measurement of $\hat{\sigma}_1$, followed by a projective measurement of $\hat{\sigma}_2$. We denote the eigenvector of those Pauli matrices by $\hat{\sigma}_j|\pm_j\rangle = \pm|\pm_j\rangle$. It is convenient to express all states in the basis that diagonalizes $\hat{\sigma}_1$.
\begin{equation}
\label{eq:basis}
|+\rangle = \alpha|+_1\rangle +\beta|-_1\rangle\hspace{2cm}|+_2\rangle = \gamma|+_1\rangle +\delta|-_1\rangle,
\end{equation}
where we consider $\alpha=\beta=\gamma=\delta=1/\sqrt{2}$ in the main text.
The KQPD can then be written as \cite{hofer:2017}
\begin{equation}
\label{eq:kqpdspins}
\begin{aligned}
\mathcal{P}_c(\Sigma_1,\Sigma_2)=&\frac{1}{(2\pi)^2}\int d\lambda_1d\lambda_2e^{i\lambda_1\Sigma_1+i\lambda_2\Sigma_2}{\rm Tr}\left\{e^{-i\frac{\lambda_2}{2}\hat{\sigma}_2}e^{-i\left(\frac{\lambda_1}{2}+\gamma_1\right)\hat{\sigma}_1}|+\rangle \langle +|e^{-i\left(\frac{\lambda_1}{2}-\gamma_1\right)\hat{\sigma}_1}e^{-i\frac{\lambda_2}{2}\hat{\sigma}_2}\right\}\\
=&\sum_{\sigma_1=0,\pm 1}\sum_{\sigma_2=\pm 1}{\mathcal{P}}(\sigma_1,\sigma_2)\delta(\Sigma_1-\sigma_1)\delta(\Sigma_2-\sigma_2),
\end{aligned}
\end{equation}
with the discrete distribution
\begin{equation}
\label{eq:kqpdspinsdisc}
\begin{aligned}
&{\mathcal{P}}(+1,+1)=|\alpha|^2|\gamma|^2,\hspace{1cm}{\mathcal{P}}(-1,+1)=|\beta|^2|\delta|^2,\hspace{1cm}{\mathcal{P}}(0,+1)=2{\rm Re}\left\{e^{-2i\gamma_1}\alpha\beta^*\gamma^*\delta\right\},\\&
{\mathcal{P}}(+1,-1)=|\alpha|^2|\delta|^2,\hspace{1cm}{\mathcal{P}}(-1,-1)=|\beta|^2|\gamma|^2,\hspace{1cm}{\mathcal{P}}(0,-1)=-2{\rm Re}\left\{e^{-2i\gamma_1}\alpha\beta^*\gamma^*\delta\right\}.
\end{aligned}
\end{equation}
Measuring $\hat{\sigma}_1$ with strength $\chi$, followed by measuring $\hat{\sigma}_2$ projectively (i.e., $\chi_2\rightarrow\infty$) results in the distribution
\begin{equation}
\label{eq:kqpdspinsmeas}
P(\sigma_1,\sigma_2|\chi)=\frac{\chi}{\sqrt{\pi}}\sum_{\sigma'_1=0,\pm 1}e^{-\chi^2(\sigma_1-\sigma_1')^2}e^{-\chi^2\delta_{\sigma'_1,0}}{\mathcal{P}}(\sigma_1',\sigma_2)|_{\gamma_1=0},
\end{equation}
where $\sigma_1$ is a continuous variable while $\sigma_2=\pm1$ is a discrete variable. 

Because the second measurement is strong, there is no need to consider characteristic functions with respect to $\sigma_2$. We thus introduce
\begin{equation}
\label{eq:charameasspin}
\begin{aligned}
\tilde{P}(\lambda_1,\sigma_2|\chi)=
\int d\sigma_1 e^{-i\lambda_1\sigma_1}P(\sigma_1,\sigma_2|\chi)=e^{-\frac{\lambda_1^2}{4\chi^2}}\sum_{\sigma'_1=0,\pm 1}e^{-i\lambda_1\sigma_1'}e^{-\chi^2\delta_{\sigma'_1,0}}{\mathcal{P}}(\sigma_1',\sigma_2)|_{\gamma_1=0}.
\end{aligned}
\end{equation}

A measurement of $\hat{\sigma}_1$ alone is described by
\begin{equation}
\label{eq:meassig1}
P(\sigma_1|\chi)=\frac{\chi}{\sqrt{\pi}}\left[|\alpha|^2e^{-\chi^2(\sigma_1-1)^2}+|\beta|^2e^{-\chi^2(\sigma_1+1)^2}\right],
\end{equation}
with the characteristic function
\begin{equation}
\label{eq:meassig1char}
\tilde{P}(\lambda_1|\chi)=e^{-\frac{\lambda_1^2}{4\chi^2}}\left[\cos(\lambda_1)+(|\beta|^2-|\alpha|^2)i\sin(\lambda_1)\right].
\end{equation}
This equation again fulfills Eq.~\eqref{eq:measimpxp}, ensuring that measurement imprecision can be isolated. We can then write
\begin{equation}
\label{eq:kspins}
K(\sigma_1,\sigma_2|\chi) = \frac{1}{2\pi}\int d\lambda_1e^{i\lambda_1\sigma_1}\tilde{P}(\lambda_1,\sigma_2|\chi)\frac{\tilde{P}(\lambda_1|\chi')}{\tilde{P}(\lambda_1|\chi)}=\frac{\chi'}{\sqrt{\pi}}\sum_{\sigma'_1=0,\pm 1}e^{-(\chi')^2(\sigma_1'-\sigma_1)^2}e^{-\chi^2\delta_{\sigma'_1,0}}{\mathcal{P}}(\sigma_1',\sigma_2)|_{\gamma_1=0}.
\end{equation}
We note that in the limit $\chi'\rightarrow\infty$, this distribution contains well separated peaks with weights that are given by $e^{-\chi^2\delta_{\sigma'_1,0}}{\mathcal{P}}(\sigma_1',\sigma_2)|_{\gamma_1=0}$.

To estimate $K$ from experimental data, we consider $N$ joint measurements, which result in outcomes $\sigma_1^j$ and $\sigma_2^j = \pm1$. We then introduce the estimate of Eq.~\eqref{eq:charameasspin} as
\begin{equation}
\label{eq:estcharjoint}
Y_{\lambda_1,\sigma_2}=\sum_{j=1}^N\delta_{\sigma_{2}^j,\sigma_2}e^{-i\lambda_1\sigma_{1}^j}.
\end{equation} 
We can then write
\begin{equation}
\label{eq:estkqpdsg}
{K}_{\rm est}=\begin{cases}
\int_{-\lambda_c}^{\lambda_c}\frac{d\lambda_1}{2\pi}e^{i\lambda_1\sigma_1}Y_{\lambda_1,\sigma_2}\frac{Y'_{\lambda_1}}{Y_{\lambda_1}}\hspace{.5cm}\text{for}\hspace{.25cm} |Y_{\lambda_1}|> c_o,\\
0 \hspace{2cm}\text{otherwise},
\end{cases}
\end{equation}
where $Y_{\lambda_1}$ and $Y_{\lambda_1}'$ denote the estimate of $\tilde{P}(\lambda_1|\chi)$ and $\tilde{P}(\lambda_1|\chi')$ respectively, following Eq.~\eqref{eq:empchar}
. We note that it is beneficial to use a $\chi'$ of moderate strength because the estimator in Eq.~\eqref{eq:estkqpdsg} becomes unreliable when the respective probability distributions vary over short scales.
\end{document}